\documentclass[manuscript]{aastex}

\newcommand{\dplan}{D_{\rm plan}}
\newcommand{\hpeb}{H_{\rm peb}}
\newcommand{\hgas}{H_{\rm gas}}
\newcommand{\mdisk}{M_{\rm disk}}
\newcommand{\mpeb}{M_{\rm peb\,acc}}
\newcommand{\mgap}{M_{\rm gap}}
\newcommand{\mrem}{M_{\rm rem}}
\newcommand{\mstar}{M_\ast}
\newcommand{\mcrit}{M_{\rm crit}}
\newcommand{\ncol}{N_{\rm col}}
\newcommand{\rpeb}{R_{\rm peb}}
\newcommand{\rfrag}{R_{\rm frag}}
\newcommand{\rdisk}{R_{\rm disk}}
\newcommand{\tplan}{t_{\rm plan}}
\newcommand{\sigmagas}{\Sigma_{\rm gas}}
\newcommand{\sigmacrit}{\Sigma_{\rm crit}}
\newcommand{\st}{\rm St}
\newcommand{\vdrift}{v_{\rm drift}}
\newcommand{\vesc}{v_{\rm vesc}}
\newcommand{\vfrag}{v_{\rm frag}}
\newcommand{\vimp}{v_{\rm imp}}
\newcommand{\vkep}{v_{\rm kep}}
\newcommand{\vrel}{v_{\rm rel}}

\shorttitle{The Diversity of Planetary Systems}

\begin{document}
\title{Pebble Accretion and the Diversity of Planetary Systems}

\author{J. E. Chambers}
\affil{Carnegie Institution for Science, 5241 Broad Branch Road NW, Washington, DC 20015}
\email{jchambers@carnegiescience.edu}

\begin{abstract}
I examine the standard model of planet formation, including pebble accretion, using numerical simulations. Planetary embryos large enough to become giant planets do not form beyond the ice line within a typical disk lifetime unless icy pebbles stick at higher speeds than in experiments using rocky pebbles. Systems like the Solar System (small inner planets, giant outer planets) can form if (i) icy pebbles are stickier than rocky pebbles, and (ii) the planetesimal formation efficiency increases with pebble size, which prevents the formation of massive terrestrial planets. Growth beyond the ice line is dominated by pebble accretion. Most growth occurs early, when the surface density of pebbles is high due to inward drift of pebbles from the outer disk. Growth is much slower after the outer disk is depleted. The outcome is sensitive to the disk radius and turbulence level, which control the lifetime and maximum size of pebbles. The outcome is sensitive to the size of the largest planetesimals since there is a threshold mass for the onset of pebble accretion. The planetesimal formation rate is unimportant provided that some large planetesimals form while pebbles remain abundant. Two outcomes are seen, depending on whether pebble accretion begins while pebbles are still abundant. Either (i) multiple gas giant planets form beyond the ice line, small planets form close to the star, and a Kuiper-belt-like disk of bodies is scattered outwards by the giant planets; or (ii) no giants form and bodies remain Earth-mass or smaller.
\end{abstract}


%
%
\section{Introduction}
The process of planet formation is still poorly understood despite recent observational and theoretical advances \citep{helled:2014, raymond:2014, winn:2015}. In particular, we do not understand precisely why planet formation in the Solar System gave rise to two distinctly different types of planet: the gas-poor inner planets, and the gas-rich, giant outer planets. Additionally, the great diversity of planetary systems seen orbiting stars other than the Sun remains to be explained.

The existence of the Sun's gas-giant planets is often ascribed to the presence of additional solid material in the colder, outer regions of the solar nebula, in the form of ices or organic compounds \citep{ida:2004, lodders:2004, kennedy:2008, dodson-robinson:2009}. This, combined with the greater gravitational reach of bodies far from the Sun, should allow solid protoplanets to grow larger in the outer Solar System, so that they can accrete massive gaseous envelopes, {\em given enough time} \citep{lissauer:1987, pollack:1996}. However, observations of other stars indicate that protoplanetary disks typically have a lifetime of only a few million years \citep{haisch:2001}. This is probably not long enough for giant planets to form if the presence of additional solid material beyond the ice line is the only factor \citep{levison:2001, inaba:2003b, thommes:2003, levison:2010}. This suggests other differences existed between the inner and outer solar nebula, which favoured the growth of large planets far from the Sun \citep{morbidelli:2015}.

The diversity of extrasolar planetary systems, and the striking differences that many show compared to the Solar System, are also surprising. To a first approximation, protoplanetary disks are likely to have much in common with one another in terms of their composition, temperature structure, and lifetime. Presumably, the same set of physical processes operate in all protoplanetary disks, both in terms of planetary growth and the evolution of the disk itself. The wide variety of planetary systems that result suggests that at least some aspects of planet formation are sensitive to small differences in the initial conditions between protoplanetary disks, or to random events that arise during planetary growth. 

In this paper, I explore whether the diversity of planetary systems, and the different characteristics of the planets in the Solar System, can be explained by the currently favored model of planet formation. In this model \citep{raymond:2014, helled:2014}, planet formation begins with dust grains embedded in a gaseous protoplanetary disk. Interactions with the gas give the dust grains a distribution of velocities, leading to frequent collisions. Initially, at least, grains stick together during collisions, forming larger aggregates \citep{blum:2000, poppe:2000}. However, pairwise growth probably stalls when the largest objects---``pebbles''---have sizes somewhere between 1mm and 1m \citep{brauer:2008, zsom:2010}. This occurs for two reasons. Firstly, collision speeds between pebbles typically increase with size due to interactions with the gas \citep{weidenschilling:1977, dullemond:2005, ormel:2007}, increasing the amount of energy that must be absorbed if a collision is to result in a merger. Secondly, gas drag causes pebbles to drift rapidly towards the star, so their lifetimes are short \citep{weidenschilling:1977}.

The next step is uncertain. A plausible scenario is that interactions with the gas concentrate large numbers of pebbles into small regions, so that their combined gravity allows them to accumulate into asteroid-sized ``planetesimals'' in a short space of time \citep{johansen:2007, cuzzi:2008}.

Planetesimals are massive enough that their gravity can pull in additional solid material, and hold on to much of this material during collisions. Previous studies suggest that gravitational interactions between planetesimals promote the rapid growth of a small subset of the population \citep{wetherill:1993, kokubo:1996}, forming planetary mass bodies dubbed ``planetary embryos''. Growth may be especially rapid if a sizeable amount of mass remains in pebbles at this stage \citep{lambrechts:2012}.  The combined effects of an embryo's gravity and aerodynamic drag acting on a pebble can lead to very large collision cross sections \citep{ormel:2010}, a process dubbed ``pebble accretion''. Finally, any embryos that reach a critical mass of at least a few times that of Earth within the lifetime of the gas disk will undergo runaway gas accretion forming gas-giant planets \citep{pollack:1996}.

Here, I study planet formation using a simplified model of planetary growth that includes all of these physical processes, and follows the evolution over the lifespan of a typical protoplanetary disk. Particular attention is paid to the main areas of uncertainty, and the sensitivity of the outcome to the initial conditions. The rest of this paper is organized as follows. Section~2 describes in more detail the model for planet formation used in the simulations. Section~3 examines the importance of how pebbles behave during collisions, and the circumstances in which they are converted into planetesimals, with particular emphasis on whether both gas-giant planets and smaller, rocky planets form in the same system. Section~4 looks at how the outcome depends on the initial sizes of the planetesimals and their formation rate, as well as the properties of the disk itself. Section~5 contains a discussion of the results, and looks at how well the simulations match the main characteristics of the Solar System and extrasolar systems. Section~6 summarizes the main findings.

%
%
\section{Simulations}
The simulations described in this paper model the orbital and collisional evolution of a population of pebbles, planetesimals and planetary embryos in an evolving protoplanetary disk. (Table~1 lists the main model parameters and corresponding symbols, together with their default values.) During collisions, particles merge, with some fraction of the total mass escaping as fragments. Pebbles and planetesimals can also move radially with respect to the star due to gas drag. At the same time, planetesimals are assumed to form directly from pebbles at a parameterized rate, although the details of this process remain unclear. I use a particle-in-a-box scheme to evolve the planetesimal mass distribution \citep{wetherill:1993}, dividing particles into bins that are logarithmically spaced in mass, with five bins per decade. The disk is divided into radial zones that are logarithmically spaced in distance from the star. Planetary embryos are treated as discrete objects that can collide or gravitationally scatter one another as well as smaller objects.

The simulations begin with $\mu$m-sized dust grains embedded in a gaseous protoplanetary disk. The surface density and temperature of the gas disk evolve over time following the analytic model described by \citet{chambers:2009}. The gas disk is assumed to have a lifetime of 3 My, which is typical for observed protoplanetary disks \citep{haisch:2001}. Dust grains are initially composed of rock and water ice, in a 1:1 mass ratio at temperatures below 150 K. The ice fraction of dust grains and pebbles declines linearly due to evaporation at temperatures above 150 K, until only the rock fraction is left at temperatures above 170 K.

Dust grains are assumed to grow into larger pebbles following a slightly modified version of the procedure used by \citet{lambrechts:2014b}. At each distance from the star, pebbles have a single radius $\rpeb$ that evolves over time due to collisions:
\begin{equation}
\frac{d\rpeb}{dt}=\frac{\Sigma_{\rm peb}\vrel}{2\hpeb\rho}\times F_{\rm frag}
\end{equation}
where $\rho$ and $\Sigma_{\rm peb}$ are the bulk density and surface density of the pebbles, and $\hpeb$ and $\vrel$ are the scale height and relative velocity of pebbles, both set by turbulence \citep{ormel:2007}, given by:
\begin{eqnarray}
\vrel&=&c_s\left(3\alpha\st\right)^{1/2} \nonumber \\
\hpeb&=&\hgas\left(\frac{\alpha}{\st}\right)^{1/2}
\end{eqnarray}
where $\st=\Omega t_{\rm stop}$ is the pebble Stokes number, $\Omega$ is the Keplerian orbital frequency, and $t_{\rm stop}$ is the stopping timescale of the pebbles due to gas drag (equal to the pebble's momentum divided by the drag force). In addition, $c_s$ and $\hgas$ are the gas sound speed and scale height, and the strength of the turbulent viscosity is parameterized by $\alpha$ using $\nu=\alpha\hgas c_s$. Pebbles are assumed to be compact objects with a bulk density of 2 g/cm$^2$, rather than fractal aggregates.

The loss of mass due to fragmentation during pebble-pebble collisions is modeled using the function $F_{\rm frag}$, given by
\begin{equation}
F_{\rm frag}=\frac{\rfrag-\rpeb}{\rfrag+\rpeb}
\end{equation}
where $\rfrag$ is the particle size at which collisions become energetic enough to be erosive rather than accretionary.

In addition to collisions, the size and surface density of pebbles at each location change over time as these quantities are advected inwards by gas drag. The drift velocity $\vdrift$ is set by two factors: the velocity difference between the particle and the gas, and also the inward motion of the gas itself:
\begin{equation}
\vdrift=-\frac{-2\eta\vkep\st}{1+\st^2}-\frac{v_{\rm gas}}{(1+\st^2)}
\end{equation}
where $v_{\rm gas}$ is the radial velocity of the gas, and $\eta$ is the fractional difference between the gas orbital speed and the local Keplerian velocity $\vkep$ \citep{birnstiel:2010}.

Pebbles are converted into planetesimals on a timescale $\tplan$, at a rate
\begin{equation}
\frac{d\Sigma_{\rm plan}}{dt}=\frac{\Sigma_{\rm peb}}{\tplan}\times F_{\rm plan}(\rpeb)
\end{equation}
where the function $F_{\rm plan}$ allows for the possibility that planetesimal formation depends on the pebble radius. Unless otherwise noted, new planetesimals are assumed to have diameters between 30 and 300 km, with total mass uniformly spaced in log mass.  We neglect the possibility that planetesimal formation yields an exponentially declining population of larger bodies, perhaps Ceres sized or even larger, as some simulations of the streaming instability have suggested \citep{johansen:2007}. If such objects do form, the subsequent evolution could be rather different than described here.

The default timescale for planetesimal formation is assumed to be $3\times 10^5$ years. This is much longer than the likely time required to form individual planetesimals, in order to be more consistent with the wide range of meteorite parent-body ages.

Planetesimals are divided into a series of mass bins, with mutual collisions and fragmentation transferring mass between bins over time. Following \citet{chambers:2014}, the number of collisions between planetesimals in bins $i$ and $j$, in a single disk zone, in time $dt$ is
\begin{equation}
\ncol=\frac{\pi R_xR_zN_iN_j\vrel\,dt}{2HA}
\end{equation}
where $N_i$ and $N_j$ are the number of planetesimals in bins $i$ and $j$, $\vrel$ is their mean relative velocity, $A$ is the area of the disk zone, and $H$ is the particle scale height, which is determined by the orbital inclination $i$ for large objects, and by turbulence for small particles \citep{youdin:2007}. Collisions between planetesimals in different zones are calculated by modifying the formula for $\ncol$ by the degree of overlap between the zones.

$R_x$ and $R_z$ are the collision capture radii in the horizontal and vertical directions in the disk. In the absence of gas drag, the capture radius would be the sum of the planetesimals' radii augmented by the effects of gravitational focussing. Gas drag modifies the capture radii in two ways. Firstly, when the gas drag timescale is comparable to or smaller than the encounter timescale, pebble-accretion effects must be taken into account. Here, I use pebble-accretion capture radii calculated by \citet{ormel:2010}, including a reduction in the collision probability in the ``hydrodynamical regime'' when pebbles are small enough to be dragged around a planetesimal by the gas flow \citep{guillot:2014}. Secondly, objects encountering large bodies can experience substantial drag within their atmospheres, enhancing the capture probability. To model this effect, I use capture radii calculated by \citet{inaba:2003a}, using the simple radiative atmosphere model described by \citet{ormel:2012}. Finally, the capture radius in the vertical direction is constrained to be no larger than the minimum scale height of the two populations involved.

The relative velocity $\vrel$ between planetesimals depends on their orbital eccentricity $e$, and to a lesser extent their inclination $i$, when these quantities are large. When $e$ and $i$ are small, the relative velocity is set by Keplerian shear---the difference in Keplerian orbital velocity $\vkep$ due to distance from the star. Here I use a simplified formula for $\vrel$ that approximately models both of these regimes:
\begin{equation}
\vrel=\vkep(e^2+h^2)^{1/2}
\end{equation}
where $h$ is the reduced Hill radius given by
\begin{equation}
h=\left(\frac{M_i+M_j}{3\mstar}\right)^{1/3}
\end{equation}
where $M_i$ is the mass of planetesimal $i$, and $\mstar$ is the stellar mass.

Planetesimal-planetesimal collisions do not lead to simple mergers in general. Instead, some fraction of the mass escapes as fragments. Here, I model fragmentation using a simplified version of the procedure described by \citet{leinhardt:2012} derived from hydrodynamic impact simulations in the regime where gravity dominates over material strength. For a collision between planetesimals with masses $M_i$ and $M_j$, I assume that the mass $\mrem$ of the largest  remnant after the collision is given by
\begin{equation}
\mrem=(M_i+M_j)\left(\frac{1+0.5\phi}{1+\phi+\phi^3}\right)
\end{equation}
where $\phi$ is the ratio of the impact energy to the gravitational strength of the combined bodies. This expression approximately matches the catastrophic and non-catastrophic disruption regimes described by \citet{leinhardt:2012} using a single formula.

I adopt a simplified version of the expression for $\phi$ used by \citet{leinhardt:2012}, which has the same dependence on the impact velocity $\vimp$, the mutual escape velocity $\vesc$, and the mass ratio $\gamma=M_j/M_i$, where $M_j\le M_i$
\begin{equation}
\phi=\frac{\gamma^2\vimp^2}{(1+\gamma)^4\vesc^2}
\end{equation}
Mass ejected as fragments is distributed into smaller mass bins following a power law with a differential exponent of -11/6. Mass that would go into fragments smaller than 100 m is assumed to become pebbles instead.

The orbital eccentricities and inclinations of planetesimals evolve due to a number of processes. Here, I calculate the evolution due to viscous stirring and dynamical friction due to gravitational encounters with other bodies, using rates calculated empirically by \citet{ohtsuki:2002}. I also calculate damping of $e$ and $i$ due to gas drag, following \citet{rafikov:2004}, and damping due to tidal interactions with the disk gas following \citet{tanaka:2004}. Finally, I include stirring of $e$ and $i$ due to density fluctuations in the gas caused by turbulence on large scales, using the rates for ideal magneto-hydrodynamics given by \citet{okuzumi:2013}.

Planetesimals with diameters $>2000$ km are promoted into embryos that are treated as discrete objects. Embryos sweep up planetesimals and pebbles within their feeding zones, and also gravitationally stir planetesimals in nearby disk zones. Gravitational encounters between embryos are treated individually and analytically using the \"Opik scheme \citep{arnold:1965}. When the minimum encounter distance is less than the sum of the embryos' radii, they are assumed to merge. Distant encounters between embryos on non-overlapping orbits are also included in an approximate way following \citet{zhou:2007}, by assuming these encounters increase the orbital eccentricities at a rate given by
\begin{equation}
\left(\frac{de^2}{dt}\right)_{\rm dist}=\left(\frac{M}{\mstar}\right)^2
\left(\frac{P_{\rm VS}}{9P}\right)
\frac{a^3}{\max([\Delta a)^3,\,12r_H^3]}
\end{equation}
where $P$ and $a$ are the orbital period and semi-major axis, $M$ is the mass of the perturbing embryo, $\Delta a$ is the orbital separation, $r_H=ha$ is the mutual Hill radius, and $P_{\rm VS}$ is given by \citet{ohtsuki:2002}.

In addition to altering the eccentricities and inclinations of nearby planetesimals, embryos can scatter planetesimals to other regions of the disk, reducing the local surface density as a result. Here, I model this process in an approximate way by calculating how many planetesimals pass within the Hill radius of each embryo in each timestep. The resulting velocity kicks received by the planetesimals are estimated using the impulse approximation, assuming that the square of the impact parameter is uniformly distributed. These velocity kicks are compared to the change in semi-major axis needed to scatter a planetesimal into various other zones in the disk. The appropriate fraction of the planetesimals are moved to these new disk zones accordingly, assuming that half of the planetesimals are scattered inwards and half outwards.

Embryos that exceed a critical mass $\mcrit$ begin to accrete gas from the disk. Here, I use a critical mass similar to that obtained by \citet{ikoma:2000}:
\begin{equation}
\mcrit=20M_\oplus\left(\frac{\dot{M}}{10^{-6}M_\oplus/{\rm y}}\right)^{1/4}
\left(\frac{\kappa}{1\ {\rm cm}^2/{\rm g}}\right)^{1/4}
\end{equation}
where $\dot{M}$ is the solid mass accretion rate of the embryo, and $\kappa$ is the atmospheric opacity, assumed here to be 0.01cm$^2$/g. For very low mass accretion rates, $\mcrit$ may become unrealistically small according to this formula. Here, I ensure that $\mcrit$ is at least 3 Earth masses, comparable to the minimum critical core mass described in simulations of gas accretion by \citet{movshovitz:2010}.

Following \citet{ida:2004}, the gas accretion rate for an embryo of mass $M$ is
\begin{equation}
\left(\frac{dM}{dt}\right)_{\rm gas}=\frac{M}{10^9\ {\rm y}}
\left(\frac{M}{M_\oplus}\right)^3
\end{equation}
The gas accretion rate is limited by the rate at which the disk can supply gas:
\begin{equation}
\left(\frac{dM}{dt}\right)_{\rm gas,\,max}=2\pi av_{\rm gas}\sigmagas
\end{equation}
where $v_{\rm gas}$ is the inward gas flow velocity, and $\sigmagas$ is the local gas surface density.

As giant planets grow, their gravity begins to clear a partial gap in the disk around their orbit. This will reduce the local gas surface density and lower the maximum gas accretion rate. \cite{crida:2006} provided an estimate of the planetary mass $\mgap$ needed to reduce $\sigmagas$ by 90\%, which I adopt here. For other planetary masses, I assume the surface density in the gap is given by
\begin{equation}
\sigmagas=\Sigma_{\rm gas,0}\left(\frac{\mgap^2}{\mgap^2+9M^2}\right)
\end{equation}
where $\Sigma_{\rm gas,0}$ is the unperturbed gas surface density. This reduction in $\sigmagas$ is assumed to extend 3 Hill radii on either side of the planet's orbit, linearly increasing to the unperturbed value at 7 Hill radii.

In addition to opening a gap in the gas disk, large planets can perturb the local gas surface density profile sufficiently to halt the inward drift of pebbles. Here I assume that planets are massive enough to do this if their mass exceeds a critical value $M_{\rm halt}$ given by \citet{lambrechts:2014a}:
\begin{equation}
M_{\rm halt}=20 M_{\rm Earth}\times\left(\frac{\hgas}{0.05a}\right)^3
\end{equation}

%
%
\section{Pebble Properties}
We begin by looking at three simulations of planetary growth in identical protoplanetary disks, but with different assumptions about pebble sticking properties and the efficiency with which pebbles are converted into planetesimals. These represent two of the main uncertainties in our current understanding of planet formation. I choose a massive disk ($0.1M_\odot$) with a large radial extent (100 AU), in order to maximize the amount of material available for planet formation, and to give pebbles a long lifespan against radial drift. Other things being equal, these choices should promote the formation of large planetary embryos capable of accreting massive gaseous envelopes. If gas giant planets fail to form in such a disk, they are unlikely to form in most other disks as well.

\subsection{Case 1}
This case begins with a 0.1 solar mass disk around a solar mass star. (Table 1 lists the main model parameters used throughout this paper except where otherwise noted.) The disk evolves viscously with $\alpha=7\times 10^{-4}$. The middle of the ice line (160 K) is initially at 2.4 AU, but moves inwards over time. Solid mass begins as $1\mu$m dust grains composed of rock above 170 K, and a 1:1 mixture of ice and rock at temperatures below 150 K, with a linear trend at intermediate temperatures. Dust grains grow via mutual collisions forming pebbles. Pebble-pebble collisions become more erosive as the collision speed increases, and collisions lead to net mass loss when the collision speed exceeds $\vfrag=1$ m/s. This value of $\vfrag$ is typical of the speed at which collisions cause fragmentation in experiments using particles composed of silica or other rocky materials \citep{guttler:2010}. Pebbles are converted into planetesimals on a timescale of $3\times 10^5$ years, independently of pebble properties.

Figure~1 shows the state of the system at four snapshots in time. The four large panels show the surface density of planetesimals in each size bin as a function of distance from the star. The different colors denote different surface densities, on a log scale, with three color bins per decade in surface density. Blue colors represent high surface densities and red colors indicate low surface densities. Discrete embryos are plotted as circles with symbol radius increasing with mass. Beneath each of the large panels is a smaller panel that shows the diameter and surface density of pebbles at each distance from the star.

At 0.02 My, the initial dust grains have grown into pebbles with sizes $\sim 1$ mm throughout much of the disk. These pebbles have already reached the size where collisions become destructive, so pebble growth stalls at this point. In the outermost parts of the disk, pebbles are still growing, due to the low rate of collisions in this region. Planetesimal formation has already begun, with planetesimals preferentially forming in the inner disk and just beyond the ice line, where the surface density of pebbles is high in each case. 

By 0.15 My, pebbles have reached the maximum size set by collisions throughout most of the disk. The maximum size declines somewhat beyond about 3 AU since the decrease in gas density with distance reduces the size for a given Stokes number, and it is the Stokes number that determines whether collisions lead to growth or net mass loss. In the inner disk, planetesimal-planetesimal collisions are starting to generate bodies larger than the initial planetesimals. There is also a small increase in the maximum planetesimal size just outside the ice line, due to the presence of extra solid material here, but the effect is minor.

At 0.5 My, the largest planetesimals exceed 2000 km in diameter, and these have been promoted to discrete embryos.  At this stage, embryos have only appeared in the region interior to 1 AU, and the maximum object size declines at larger distances, with a small jump at the ice line. The pebble surface density has declined somewhat due to a combination of radial drift and planetesimal formation. However, drift rates are modest due to the small size of the pebbles, and pebbles remain abundant in the outer disk at this stage.

After 3 My, the largest bodies have reached the mass of Mercury, and this size is roughly independent of distance out to about 3.5 AU. This is a typical outcome of ``oligarchic growth'', in which gravitational perturbations from large embryos tend to stir up the relative velocities of nearby planetesimals more than smaller embryos do, leading to a negative feedback on the growth rate \citep{kokubo:1998}. Pebble accretion modifies this picture somewhat, but the small size of the pebbles in this case means that pebble accretion is relatively inefficient, both because the capture probability is low, and the scale height of the pebbles is large. As a result, pebbles have typically contributed only a few percent of the total mass of these embryos. Growth is negligible beyond about 7 AU, due to the very low rate of planetesimal-planetesimal collisions in this region. The largest bodies here are only slightly more massive than the largest planetesimals formed directly from pebbles.

\subsection{Case 2}
The difficulty of forming large embryos more than a few AU from a star has been known for some time \citep{lissauer:1987, levison:2001, thommes:2003}, and this was an important motivation for developing models that include pebble accretion. In the previous section, we saw that using pebbles that fragment at 1 m/s is not sufficient to form embryos large enough to become giant planet cores by 3 My because the pebbles are too small for pebble accretion to be effective. The marked differences between the Sun's inner and outer planets has led \citet{morbidelli:2015} to suggest that pebbles might have had different sizes in the inner and outer solar nebula, in a way that preferentially promoted efficient planetary growth beyond the ice line. There is some theoretical and experimental evidence to suggest that ice-rich pebbles will grow larger than pebbles made of rock alone due to differences in their sticking properties \citep{supulver:1997, wada:2009} \citep{gundlach:2015} (but see also \citet{hill:2015}). If true, the presence of larger pebbles beyond the ice line would increase the importance of pebble accretion, potentially allowing more massive embryos to form. I test this possibility in this section. 

Case 2 begins with identical initial conditions and model parameters to Case 1, except that when the temperature $T<150$ K, the pebble fragmentation speed $\vfrag$ is 3 m/s instead of 1 m/s. When $T>170$ K, $\vfrag=1$ m/s as before, with a linear interpolation in the pebble fragmentation speed at intermediate temperatures. Since the ice component of pebbles begins to evaporate when $T>150$ K, pebbles will become weaker or less sticky as the drift inwards across the ice line.

Figure~2 shows the evolution at 4 snapshots in time in this case. Several differences from Case 1 are apparent. In Case~2, pebbles grow larger beyond the ice line, as expected. The pebble diameter increases by roughly an order of magnitude at the ice line, from about 1.5 mm to 1.5 cm, which increases the efficiency of pebble accretion, and also increases the inward drift rate of the pebbles. Once pebbles drift inwards across the ice line, their diameter decreases, and their drift rate slows. This leads to a pile up of mass in the inner disk, and more efficient planetesimal formation and growth in this region compared to Case~1.  The effect of this mass influx can be seen clearly in the higher pebble surface densities from 0.15 My onwards in the inner disk in Case~2 compared to Case~1.

Beyond the ice line, pebbles are swept up much faster in Case~2 than in Case~1, leading to the formation of large planetesimals and embryos out to about 5 AU by 0.5 My. By 3 My, the largest body at 2.9 AU has a mass roughly 6 times that of Earth. Several other objects larger than Earth have formed, both inside and outside the ice line. These bodies are much larger than those at the same stage in Case~1, which can be attributed mainly to the large influx of mass into the inner disk, and the increased efficiency of pebble accretion beyond the ice line. By 3 My, pebbles have typically contributed  50--80\% of the mass of the largest embryos outside the ice line in Case 2, compared to about 10--20\% for embryos inside the ice line, and a few percent for embryos throughout the disk in Case~1. 

As in Case~1, the masses of the largest embryos at 3 My vary only slightly with distance from the star over an extended region of the disk. However, this region now extends to about 7 AU rather than 3.5 AU. The largest embryos inside the ice line have swept up most of their rivals by this stage, and are depleting the population of planetesimals too. Beyond the ice line, many embryos are still present on overlapping orbits, and oligarchic growth is still at a relatively early stage. Growth rates beyond about 8 AU are very slow as they were in Case~1.

By 3 My, most of the pebbles have been removed by a combination of inward drift, planetesimal formation, and accretion onto larger bodies. The surface density of pebbles is so low at this point that the pebble-pebble collision timescale becomes long compared to the inward drift timescale. As a result, inward advection of pebbles blurs the discontinuity in pebble size that had existed at the ice line at earlier times.

Some of the differences between Cases 1 and 2 can be seen more easily in Figure~3, which shows the growth of one of the largest bodies in each simulation. These objects are located just outside the ice line, at about 3 AU in each case. Prior to the formation of discrete embryos, the figure shows the mass of the largest planetesimals at the same location. Also shown is the local surface density of pebbles versus time.

At early times, the surface density of pebbles at 3 AU is somewhat higher in Case~2 than Case~1 due to the effects of inward drift. However, this modest difference is not enough to explain the great difference in growth rates for the largest objects near 3 AU in each case. Instead, this can be attributed to the roughly order of magnitude difference in pebble size. The larger pebbles in Case~2 are swept up at a more rapid rate than those in Case~1, both because the capture radius is larger for the larger pebble size \citep{ormel:2010}, and because the scale height is smaller for these pebbles, leading to a greater space density in the vicinity of a growing embryo.

The growth rate slows markedly for the embryo in Case~2 after about 0.3 My, due to the steep decline in pebble surface density. This decline is mostly due to the decreased flux of pebbles arriving from the outer disk, rather than the sweep up of pebbles by the embryos themselves.  The surface density of pebbles at 3 AU also declines in Case~1, but at a much slower rate. The pebbles arriving from the outer disk are smaller in this case, and drift inwards more slowly, so there is still a substantial reservoir of pebbles in the outer disk after 0.3 My in Case~1. In fact, the slow decline in pebble surface density at 3 AU is largely offset by the increasing efficiency with which the largest body sweeps up these pebbles, so this body actually begins to close the gap in mass with the corresponding body in Case 2. By 3 My, however, this gap still amounts to almost 2 orders of magnitude in mass.

\subsection{Case 3}
The outcome of Case~2 represents a substantial improvement over Case~1 in terms of reproducing the characteristics of the Solar System, but the match is still rather poor. For example, the largest embryos at 3 My are probably still too small to undergo runaway gas accretion and form planets similar to Jupiter, unless the gas disk survives for another few My. More importantly, the largest bodies in the region occupied by the terrestrial planets are already more massive than Earth, and the total solid mass interior to 1.5 AU is about 10 Earth masses---5 times larger than in the Solar System. It seems unlikely that this mass will be reduced substantially by subsequent events, and it may increase further.

The shortcomings of Case~2 suggest that we are still missing an important factor in the formation of the Sun's planetary system. The presence of large pebbles beyond the ice line promotes the growth of large embryos in this region, which may become giant planets, but it also leads to a large flux of mass into the inner disk, leading to the formation of terrestrial planets more massive than those we see in the Solar System. 

In this section, I examine one modification that can resolve this problem: I assume that the rate of planetesimal formation in a given location depends on the local properties of the pebbles, specifically their size. Such a scenario is supported by simulations of planetesimal formation from pebbles via the streaming instability. For example, \citet{carrera:2015} find that the local solid-to-gas ratio in the disk needed to initiate efficient planetesimal formation varies with the pebble size, and thus their Stokes number. Pebbles with Stokes numbers $\st\sim0.1$ can form planetesimals for solid-to-gas ratios similar to the solar metallicity (assuming all the solid mass is in pebbles), while higher solid-to-gas ratios are required for pebbles with larger or smaller values of $\st$.

Here, I adopt a model in which the rate at which pebbles are converted into planetesimals depends on the local pebble-to-gas surface density ratio and the pebble Stokes number $\st$. The planetesimal formation rate is given by
\begin{equation}
\frac{d\Sigma_{\rm plan}}{dt}=\frac{\Sigma_{\rm peb}}{t_{\rm plan}}
\left(\frac{\Sigma_{\rm peb}}{\Sigma_{\rm peb}+\sigmacrit}\right)^2
\label{eq_dsigmadt_case3}
\end{equation}
where the form of $\sigmacrit$ is an empirical fit to Fig~8 of \citet{carrera:2015}:
\begin{equation}
\sigmacrit=\left(\frac{\Sigma_{\rm gas}}{100}\right)
10^{\theta^2/4}
\end{equation}
where
\begin{equation}
\theta=\log_{\rm 10}(10\,\st)
\end{equation}
Thus I assume that the rate of planetesimal formation falls off steeply when $\Sigma_{\rm peb}<\sigmacrit$, but some formation continues to take place. In all other respects, the simulation in Case~3 uses an identical model and initial conditions to Case~2. In particular, the pebble fragmentation speed is higher for ice-rich pebbles than for rocky pebbles.

Figure~4 shows the evolution in Case~3 at four snapshots in time. Clear differences with Case 2 are apparent from an early stage. In Case~2, large numbers of planetesimals have formed throughout the disk by 0.02 My, whereas in Case~3, planetesimal formation is restricted to a region spanning several AU just beyond the ice line. Pebbles inside the ice line have Stokes numbers that are smaller by an order of magnitude, so planetesimal formation proceeds much more slowly here than outside the ice line. Pebbles in the outer regions of the disk are still growing, so planetesimal formation is inefficient in this region as well.

Because efficient planetesimal formation in Case~3 is restricted to a narrow region outside the ice line, this means that subsequent growth proceeds much more rapidly here than elsewhere in the disk. This can be seen in Figure~4 at 0.15 My where the largest bodies between about 2.5 and 4.5 AU are roughly 1500 km in diameter, compared to less than 500 km in most of the rest of the disk. 

By 0.5 My, the largest bodies in Case 3 have grown to about 2 Earth masses, and these are all located just inside the ice line or up to several AU beyond it. Embryos are much smaller interior to 1.5 AU. In the inner disk, there is a marked trend of decreasing embryo mass with increasing distance from the star. This trend is mostly due to the decreasing rate of planetesimal-planetesimal collisions with increasing distance from the star, due to their lower space density. The effectiveness of pebble accretion can also decrease with distance from the star in this region \citep{levison:2015}. However, at this stage pebbles have contributed only 10--20\% of the mass of embryos interior to the ice line, so this effect is relatively minor here. 

It is interesting to note that the surface density of pebbles inside the ice line is roughly a factor of 3 lower at 0.5 My in Case~3 than in Case~2. This is despite the fact that the conversion of pebbles into planetesimals is slower in this region in Case~3, and the rate at which pebbles are swept up by embryos is also lower in Case~3 because the embryos in this region are smaller than those in Case~2. This suggests that the inner parts of the disk are being starved of pebbles in Case~3. Many of the pebbles drifting inwards from the outer disk are swept up by the large embryos between 2 and 7 AU instead of reaching the inner disk. 

Figure~5 shows this more clearly. The figure shows the surface density of pebbles versus distance from the star at 0.5 My in Cases~2 and 3. In the outer disk, the surface density of pebbles is somewhat higher in Case 3, reflecting the inefficiency of planetesimal formation in this region earlier in the simulation. The pebble size as a function of distance is almost identical in the two simulations at 0.5 My, and the gas disk is the same in each case. As a result, the inward flux of pebbles is higher in Case 3 than in Case 2. However, inside 5 AU, the pebble surface density in Case~3 starts to decline as we move closer to the star, deviating from the trend seen in Case~2. The surface density of pebbles in Case~3 falls below that in Case~2 everywhere inside 4 AU, despite the larger inward flux from the outer disk. Inside the ice line, the pebble surface density in Case~3 follows the same trend as Case~2 once more, albeit with values a factor of 3--4 lower. This indicates that the surface density of pebbles in the inner disk is substantially modified by the efficiency of pebble accretion by embryos immediately beyond the ice line. These embryos are responsible for the low pebble surface densities in the inner disk in Case~3.

One other process can be seen in the outer regions of the disk at 0.5 My in Figure~4. Several embryos are present beyond 8 AU at this point even though planetesimal growth is minimal in this region. These embryos all formed closer to the star and were scattered outwards by encounters with other embryos. Their orbital eccentricities were later damped by dynamical friction and disk tides, leaving the embryos stranded in the outer disk. In addition, other embryos were scattered outwards onto unbound orbits and lost. This behaviour is more marked at 3 My in Case~3, where a belt of embryos somewhat analogous to the primordial Kuiper belt has formed via outward scattering. This shows that planet formation can be distinctly non-local in nature, even while the protoplanetary disk is still present.

By 3 My, the situation in Case~3 is very different than Case 2. Six embryos have grown massive enough to undergo rapid gas accretion, forming a set of gas giant planets on narrowly spaced, nearly circular orbits. The recipe for gas accretion and gap clearing used here means that these planets have comparable masses, increasing modestly with distance from the star, from 170 to 470 Earth masses. This configuration is likely to be unstable on long timescales, so the final planetary system formed in this case will look different. However, it is clear that multiple gas giant planets can form within a few My in this model for planet formation.

In addition, this simulation shows that the differences between the Sun's inner and outer planets can be reproduced, albeit qualitatively. The two innermost planets at 3 My in Figure~4 have masses of 0.02 and 1.4 Earth masses, representing the great majority of mass in the innermost disk. The contrast between these objects and the gas-giant planets located further from the star bears some resemblance to the Solar System. Some differences are obvious however. In particular, the region containing the giant planets lies closer to the star than in the Solar System---the innermost giant planet in Figure~4 is just outside 1 AU. This may be a reflection of the model used here for the evolution of the protoplanetary disk. A hotter disk, with an ice line further from the star, might yield a better match with the Sun's planets by leaving more room for terrestrial planets to form.

%
%
\section{Planetesimal and Disk Properties}
In this section, I examine some other factors that can affect the outcome of planet formation: the characteristics of the planetesimals formed from pebbles, and the properties of the protoplanetary disk. 

\subsection{Planetesimal Size}
One of several important unknowns in models for planet formation is the initial size of the planetesimals that form the building blocks for subsequent growth. There is some evidence that planetesimals in the Solar System were typically large---perhaps 100 km in diameter---but this is still a matter of debate \citep{morbidelli:2009, weidenschilling:2011}. It makes sense to consider other possible planetesimal sizes until this issue is resolved. In this section, I examine the role of planetesimal size for planet formation.

Figure~6 shows the outcome of four simulations each using different sizes for  planetesimals when they first form from pebbles. In each case, the planetesimals have a range of sizes spanning an order of magnitude in diameter, but the minimum and maximum values are different for each simulation, as indicated in the figure. The other model parameters are the same as those used in Case~3 in Section~3. Icy pebbles have a higher fragmentation speed (3 m/s) than rocky pebbles (1 m/s), and the efficiency of planetesimal formation depends on the pebble Stokes number following Eqn.~\ref{eq_dsigmadt_case3}

The four large panels in Figure 6 show the state of the surviving planetesimals and embryos after 3 My, while the small panels show the remaining pebbles in each case. It is clear that the outcomes can vary substantially depending on the planetesimal size used. In each of the runs using the two largest planetesimal sizes, multiple gas giant planets have formed by 3 My. A couple of small, rocky planets orbit close to the star in each case, and a belt of embryos and planetesimals remains in the outer disk. These two simulations actually give remarkably similar results.

The other two cases are very different. In the runs using the two smallest planetesimal sizes, no embryos have grown large enough to accrete gas by 3 My, and so no giant planets are present. Growth is restricted to a much narrower region of the disk than the previous cases, with no embryos or large planetesimals beyond 5 AU and 2.5 AU for the cases with planetesimal diameters $\dplan=25$--250 km and 10--100 km respectively. A substantial amount of mass remains in planetesimals across most of the disk in each case, so further growth can take place. However, few pebbles remain at this time, so pebble accretion will not play a significant role if the disk survives longer than 3 My.

The differences between the second and third cases in Figure~6, with $\dplan=30$--300 and 25--250 km, are striking, even though the initial planetesimal sizes differ only slightly. This suggests a threshhold effect is at work, and that the outcome depends sensitively on whether this threshhold is reached. 

This becomes clearer in Figure~7. The upper panel in the figure shows the growth of the first large object to appear beyond the ice line in three of the four simulations in Figure~6 together with an additional case with $\dplan=20$--200 km. The large objects are located between about 2.8 and 3.6 AU, depending on the simulation. The lower panel in the figure shows the surface density of pebbles at the same location in the disk.

In Figure~7, the simulation with the largest planetesimals ($\dplan=30$--300 km) passes through several evolutionary stages. The large object shown in the upper panel undergoes a period of rapid growth early in the simulation beginning when its mass slightly exceeds $10^{-5}$ Earth masses. At this stage, the surface density of pebbles is high, and increasing slowly as more pebbles drift into the region from the outer disk. Growth of the big body slows markedly after 0.3 My, which coincides with a steep drop in the local pebble surface density as the population of pebbles in the outer disk becomes depleted. The large body grows slowly until about 0.7 My, when it becomes massive enough to accrete gas, and a second epoch of rapid growth occurs. The pebble surface density drops very rapidly at this point as the large body is now massive enough to prevent the inward drift of additional pebbles \citep{lambrechts:2014a}.

Pebble accretion becomes important when an embryo's gravity is strong enough to produce a large deflection in a pebble's trajectory on a timescale shorter than the stopping time of the pebble due to gas drag \citep{ormel:2010}. Otherwise, the pebble remains too strongly coupled to the gas for the embryo's gravity to be important. Since the strength of an embryos' gravity increases with its mass, there comes a point when pebble accretion becomes important, and growth can be rapid after this threshhold is crossed. 

Previous studies have found a similar period of rapid growth due to pebble accretion when the mass of an embryo exceeds a particular threshhold \citep{lambrechts:2012, chambers:2014, kretke:2014}. This mass depends on the Stokes number of pebbles involved, and thus their size. Following \citet{ormel:2010}, pebble accretion becomes important when the mass of a planetesimal exceeds $\mpeb$, given by
\begin{equation}
\mpeb=\frac{\eta^3\st}{4}M_\ast
\end{equation}
where $\eta\vkep$ is the velocity of pebbles with respect to the local Keplerian velocity, resulting from gas drag (this formula is valid for $\st<1$, and assumes that the planetesimal's eccentricity $e<\eta$). In the simulations shown in Figure~7, $\eta\sim 0.002$ just beyond the ice line, and the pebbles in this region have $\st\sim 0.01$, so the onset of pebble accretion occurs at $\mpeb\sim 10^{-5}$ Earth masses.

The other cases in Figure~7 pass through some of the same stages to varying degrees.
In the simulations with $\dplan=25$--250 and 20--200 km, rapid growth due to pebble accretion also occurs when the largest object's mass exceeds about $10^{-5}$ Earth masses. However, due to the lower initial masses of the planetesimals, some time is required to reach this threshhold mass, and so the onset of rapid growth is delayed. Rapid growth ceases when the supply of pebbles is truncated. This timescale depends mainly on the radial drift rates of the pebbles and the radial extent of the disk, and is about 0.3 My for the cases shown here, independent of planetesimal size.

When the initial planetesimal size is decreased, the era of rapid growth due to pebble accretion becomes shorter, and the mass of the largest body at 0.3 My is reduced. When $\dplan$ is less than 30--300 km, the era of rapid pebble accretion is too short to produce embryos large enough to accrete gas within the remaining lifetime of the disk, and giant planets fail to form. In the smallest case considered in Figure~7, the largest planetesimals don't even become massive enough to undergo efficient pebble accretion before the supply of pebbles diminishes. 

The outcome of each simulation depends to a large extent on two timescales: the time required to reach the threshhold for the onset of rapid pebble accretion, and the lifetime of pebbles within the disk due to radial drift. If objects exceeding the threshhold for pebble accretion appear at an early stage, while pebbles are still abundant, they can grow massive enough to accrete gas and form giant planets. Otherwise, these objects will remain small.

One factor we haven't considered so far is whether the population of pebbles in the inner parts of the disk can be replenished in other ways apart from inward drift from the outer disk. In fact, pebbles are continuously regenerated in the simulations as a result of disruptive collisions between planetesimals. This plays a role at late times, but only a minor one. 

For example, the dotted line in the lower panel of Figure~7 shows the surface density of pebbles just outside the ice line in a simulation in which planetesimals are prevented from growing after they form. The only processes operating in this case are the growth and inward drift of pebbles, and the conversion of pebbles into planetesimals. For the first million years or so, the surface density of pebbles in this case closely tracks the simulations with growth included, suggesting that radial drift is the main factor controlling the abundance of pebbles. (The simulation with $\dplan=30$--300 km, in which a gas giant forms, is an exception since this object is large enough to prevent radial drift of pebbles). At later times, the surface density of pebbles is higher in the simulations that include growth, and this must be due to the regeneration of pebbles by planetesimal collisions. However, the surface density of pebbles at this stage is several orders of magnitude smaller than that in the first 0.3 My, which suggests that the regeneration of pebbles via collisions is unlikely to lead to the kind of rapid growth needed for embryos to form giant planets.

\subsection{Planetesimal Formation Time}
A second aspect of planetesimal formation that remains unclear is the timescale on which planetesimals appear in a protoplanetary disk. Studies of planet formation often begin with all the solid material in the disk already present in planetesimals. However, the range of ages deduced for the parent bodies of meteorites suggests that planetesimal formation in the solar nebula continued for an extended period of time, perhaps several million years \citep{kita:2013}.

Figure~8 shows the outcome of four simulations with different timescales $\tplan$ for planetesimal formation, ranging from $10^5$ to $3\times 10^6$ years. These values are likely to bracket the true timescales for planetesimal formation in the solar nebula. All other parameters are the same as Case~3 in Section~3, with $\dplan=30$--300 km. Each panel in the figure shows the state of a simulation after 3 My. All four simulations ended with multiple gas-giant planets, and a handful of small, rocky planets close to the star, together with a belt of embryos and planetesimals in the outer disk. The main differences are the number of giant planets that form, and the radial extent of the region containing the giant planets. However, these quantities don't depend monotonically on the planetesimal formation time.

The results suggest that the outcome of planetary growth doesn't depend strongly on $\tplan$, unlike the sensitive dependence on planetesimal size that we saw in Section~4.1. This conclusion is reinforced by Figure~9, which shows the growth history of the first large body to form outside the ice line in each of the simulations shown in Figure~8, together with the local pebble surface density in each case. The growth curves in the upper panel of Figure~9 are all similar, the main difference being a modest variation in the time when gas accretion begins. The evolution of the pebble surface density, shown in the lower panel of Figure~9, is also similar for the 4 cases, at least until large planets form that are able to truncate the radial flow of pebbles from the outer disk.

It may seem surprising that the timescale for planetesimal formation has so little effect on the outcome. However, the result makes sense if one considers that pebble accretion is the dominant source of growth for bodies beyond the ice line, provided that their mass exceeds about $10^{-5}$ Earth masses. This is only slightly larger than the most massive planetesimals ($4.7\times 10^{-6}$ Earth masses) that form directly from pebbles. Thus, rapid growth can take place independently of $\tplan$ as long as at least a few large planetesimals form early in the simulation. Even for the longest timescale considered here, $\tplan=3\times 10^6$ years, some large planetesimals form within the first few tens of thousands of years. These objects can sweep up pebbles while they remain abundant, eventually forming embryos massive enough to accrete gas before the disk disperses.

There is some suggestion in Figures 8 and 9 that slow planetesimal formation actually favors the early appearance of multiple giant planets, at least for $\tplan\le 1$ My. Presumably, this is because as long as some large planetesimals form, it is more efficient for the subsequent growth of these objects if most of the solid mass remains in the form of pebbles rather than being converted into planetesimals. Of the four cases shown in Figure~9, a giant planet first appears beyond the ice line in the run with $\tplan=1$ My, beating the cases with smaller $\tplan$ by a small margin. When $\tplan=3$ My, however, giant planets take longer to appear (and only three form by 3 My compared to six when $\tplan=1$ My), which suggests that the optimum planetesimal formation time is less than 3 My, at least for the parameters used here.

In the innermost region of the disk, where pebble accretion is less important than beyond the ice line, we might expect the planetesimal formation timescale to play a greater role. To some extent this is the case. The region containing planets that are too small have accreted massive gaseous envelopes is wider when $\tplan=10^5$ and $3\times 10^6$ years than for intermediate values of $\tplan$, and the number of rocky planets at 3 My is correspondingly larger. However, the differences are still relatively modest, which suggests that the evolution in the terrestrial-planet region is shaped to a large extent by processes happening further out in the disk.

\subsection{Disk Radius and Viscosity}
Observations indicate that protoplanetary disks have a range of radii \citep{vicente:2005, andrews:2010}. This is likely to be important for planet formation since the lifetime of pebbles with respect to radial drift depends on the size of the disk, in addition to the size of the pebbles. Other things being equal, large disks may retain an abundant population of pebbles for longer than small disks, potentially favoring the growth of giant planets by pebble accretion. The strength of turbulence in the disk is also important since it determines the relative collision velocity of pebbles and hence the size at which they start to fragment. Strong turbulence will also increase the scale height of pebbles within the disk, lowering their space density as a result. In this section, I investigate the effect of varying the disk radius $\rdisk$ and turbulence strength $\alpha$ within the model used here.

Figure~10 shows the final state of four simulations using different values of $\rdisk$ and $\alpha$ at 3 My. All other model parameters are the same as those used in Case~3, and the results of these simulations can be compared with Case~3 in the last panel of Figure~4. 

The upper left panel of Figure~10 shows a simulation with $\alpha=5\times 10^{-4}$, somewhat smaller than the turbulence strength used in Case~3. The outcome in these two cases are similar: multiple gas-giant planets have formed (six in Case~3, eight in the simulation in Figure~10), and a pair of small, rocky planets orbit close to the star in both cases, with the larger rocky planet comparable in mass to Earth. Beyond the outermost giant planet is a belt of planetesimals and embryos that are still evolving. 

The upper right panel of Figure~10 shows a run with $\alpha=10^{-3}$, slightly larger than Case~3. In this case, the outcome is very different. No giant planets have formed within 3 My, and the largest embryos are only about 0.5 Earth masses. Inside the ice line, the largest bodies have masses comparable to the Moon. A considerable amount of mass remains in unaccreted planetesimals throughout the disk, although most of the pebbles have gone.

The lower left and lower right panels of Figure~10 show simulations with $\rdisk=50$ and 150 AU, compared to 100 AU in Case~3. In all three cases, $\alpha=7\times 10^{-4}$. The differences here are even more striking than those involving different turbulence strengths. When $\rdisk=50$ AU, almost no growth takes place beyond the ice line, while seven gas-giant planets are able to form when $\rdisk=150$ AU. I note that the total disk mass is the same in both cases, so that the initial surface densities are actually lower in the run with larger disk radius. Naively, one would expect greater growth in the case with higher initial surface density, but this is not the case due to the great mobility of pebbles within the disk.

The simulations in Figure~10 show the same dichotomy of outcomes as those in Figure~6 where the varying factor was the initial planetesimal size. This suggests that the same threshhold effect is at work when varying the turbulence strength or the disk radius. In all three cases, the outcome depends on whether large planetesimals can undergo efficient pebble accretion during a temporary episode early in the disk's history that is marked by high pebble surface densities. Pebble accretion is more efficient when the pebbles are large (corresponding to low values of $\alpha$), or the planetesimals are large (large values of $\dplan$), or the pebbles survive for a longer time (which occurs when $\rdisk$ is large). In these circumstances, giant planets are likely to form. If these criteria are not met, pebble accretion is much less effective, and giant planets do not form.

With these comments in mind, Figure~11 shows the largest planet to form within 3 My for simulations with a range of values of $\rdisk$ and $\alpha$. The phase space is clearly divided into two regions. Towards the lower right of the figure (large $\rdisk$ and small $\alpha$), giant planets roughly similar in mass to Jupiter form in every case. Towards the upper left of the figure (small $\rdisk$ and large $\alpha$), no giant planets form, and the largest planets have masses comparable to or smaller than Earth. The boundary between the two regions in Figure~11 is sharp. A small change in $\rdisk$ or especially in $\alpha$ is sufficient to turn a system that would otherwise form only Earth-mass planets into a system with giant planets similar to Jupiter.

\subsection{Turbulent Stirring}
In the simulations described above, the orbital eccentricities and inclinations of planetesimals and embryos are excited by density fluctuations in the gas caused by turbulence. This excitation increases the relative velocities of bodies, reducing gravitational focussing and slowing growth as a result. The adopted stirring rates are those given by \citet{okuzumi:2013} for ideal magneto-hydrodynamics. It is possible, however, that the degree of excitation in real protoplanetary disks is different than assumed here, or absent altogether.

To examine the importance of this factor, I reran the simulation described in Case~3 above (shown in Figure~4) without including excitation due to turbulent density fluctuations. Growth rates were slower, as expected, but the difference is relatively small. Typically, planetesimals and small embryos took 10--20\% less time to reach a given mass when turbulent stirring was neglected compared to Case~3. Thus turbulent stirring played only a minor role in determining growth rates. The strength of turbulent stirring depends on the disk viscosity parameter, $\alpha$ which is quite small in this case. Turbulent stirring is likely to be more important for disks with stronger turbulence than used here.

%
%
\section{Discussion}
In the standard model of planet formation, two kinds of planets are predicted to form \citep{helled:2014, raymond:2014}. Close to a star, the condensible mass in a protoplanetary disk is limited to rocky materials, and the gravitational reach of protoplanets is small due to their proximity of the star. This leads to the formation of small, terrestrial planets. Further away from the star, icy materials can condense, and the gravitational reach of protoplanets is larger, leading to the formation of more massive bodies that can accrete gas from the disk and become gas giant planets.

In this paper, I have explored the standard model with the addition of pebble accretion dynamics---the increased capture radius of a protoplanet with respect to small particles that undergo strong gas drag \citep{ormel:2010}. Previous studies have found that the formation of gas giant planets is challenging within the lifetime of a typical disk in the absence of pebble accretion, and pebble accretion has been proposed as a possible solution \citep{lambrechts:2012}. Here, I find that even with pebble accretion, giant planet formation is unlikely if pebbles have properties similar to those found in lab experiments for silica particles. Such pebbles are likely to remain too small for pebble accretion to be effective.

Two modifications change the picture completely. When these are included, the simulations described in this paper typically yield both terrestrial and giant planets in systems that somewhat resemble the Solar System. Firstly, it has been suggested that ice-rich pebbles should stick at higher collision speeds than rocky pebbles \citep{supulver:1997, wada:2009}. This will allow pebbles to grow larger beyond the ice line, aiding the formation of giant planets here \citep{morbidelli:2015}. I find that increasing the maximum sticking speed for icy pebbles by a factor of 3 allows pebbles to grow 10 times larger beyond the ice line. This greatly increases the effectiveness of pebble accretion, allowing planetary embryos several times more massive than Earth to grow in a few My. Some of these could grow to become giant planets.

However, this leads to a second problem. The presence of large pebbles beyond the ice line means that these pebbles drift inwards rapidly, depositing large amounts of mass in the inner regions---much more than we see in the terrestrial planets of the Solar System. A second modification overcomes this problem. Following studies of planetesimal formation via the streaming instability \citep{carrera:2015}, I assume that planetesimals form more efficiently with increasing pebble size. When this effect is included, planetesimals form and grow rapidly outside the ice line, forming multiple gas-giant planets in a few My. Conversely, planetesimal formation is inefficient inside the ice line. This, coupled with the low efficiency of pebble accretion in the inner region means that the terrestrial planets remain small.

The formation of giant planets depends on an early phase in which large pebbles remain abundant in the region inside 5 AU due to a high flux of pebbles drifting inwards from the outer disk. If the pebbles are large, they are lost rapidly from a given region via inward radial drift, but they are immediately replaced by new pebbles drifting into the region from larger distances. This process continues until the great majority of mass in the outer disk has been lost, at which point the pebble surface density falls rapidly everywhere. The duration of the phase of abundant pebbles depends on the drift rate and thus their size. However, the surface density of pebbles at a particular location is {\em not\/} strongly dependent on pebble size. In a typical simulation, the surface density of pebbles inside 5 AU increases slowly for several hundred thousand years, and then drops rapidly. The outcome of planetary growth beyond the ice line depends on a race between the inward drift of these pebbles, and the ability of planetesimals to sweep up the pebbles before they are gone. For giant planets to form, the growth of their cores must be largely complete by the time this phase ends. 

Several factors may lengthen the lifetime of pebbles. They could be smaller than those assumed here, leading to lower drift rates, or the disk radius could be larger. However, both these changes would also negatively affect planetary growth rates as well. New pebbles can form via planetesimal-planetesimal collisions, but I find that the resulting pebble surface density is orders of magnitude lower than at early times, so this process is unlikely to alter the outcome significantly. Average pebble drift rates may be lower than assumed here if drift is not always inwards. For example, long-lived vortices or local pressure maxima in the disk would slow the loss of pebbles \citep{haghighipour:2005, kretke:2007, richard:2013}.

The importance of an early episode of rapid growth in the presence of a short-lived population of pebbles suggests that the initial structure and early evolution of a protoplanetary disk could be very important. Disks are thought to be compact initially, and expand radially over time due to viscous dissipation and accretion of material from the star's surrounding molecular cloud core \citep{birnstiel:2010}. If early outflow is protracted, and pebbles remain small, they can be carried away from the star, allowing them to survive until large planetesimals form. However, if the pebbles are too large, their inward drift due to the headwind they experience will exceed the rate at which outflowing gas can carry them outwards, and the pebble lifetimes will remain short. An important caveat here is that an early compact phase is likely to be associated with high temperatures in the disk, causing drifting pebbles to evaporate before they reach the inner edge of the disk. This, coupled with diffusion of the resulting vapor across the disk may lead to substantial spatial and temporal variations in the composition of pebbles that are not considered here \citep{ciesla:2006, stevenson:1988}.

In the simulations presented here, the terrestrial planets mostly grow from planetesimal-planetesimal collisions, with pebbles typically contributing 10--20\% of their final mass. Inside the ice line, growth rates decrease rapidly with distance from the star, leading to progressively smaller planetary embryos with distance. A similar mass-distance trend was found in simulations of terrestrial-planet formation with pebble accretion by \citet{levison:2015}, and these authors suggested it may explain the low mass of Mars compared to Earth and Venus. However, these authors assumed that pebbles were much larger in the terrestrial-planet region than those used in this study. As a result, the growth of terrestrial planets was dominated by pebble accretion rather than planetesimals, which is the case in the simulations in this paper. I also note that the smooth trend of declining embryo masses with distance inside the ice line is generally disrupted once gas giants form beyond the ice line----a process not  included by \citet{levison:2015}. For this reason, I believe the origin of Mars's small mass warrants further study.

\citet{morbidelli:2015} described a similar scenario to the one presented here. The formation of giant planets in the outer Solar System, and their absence  in the inner Solar System is attributed to the presence of larger pebbles beyond the ice line and the importance of pebble accretion in controlling the outcome of planetary growth. These authors showed that individual planetary embryos can grow much larger beyond the ice line but they did not consider the behavior of the system as a whole. As we saw in Case~2 in Section 3 (Figure 2), this situation is likely to lead to excessive mass in the terrestrial planet region compared to the Solar System. Overcoming this problem probably requires an additional process such as a difference in planetesimal formation efficiency on opposite sides of the ice line due to pebble size. Case~3, shown in Figure 4, shows that such a modification produces systems more like our own.

The results of this study can be compared with simulations of planetary growth by \citet{chambers:2008} that did not include pebble accretion. In that study, solid material begin in the form of planetesimals and $10^{-4}$-Earth-mass embryos, but the physical processes studied were similar apart from pebble accretion. Using 200-km-diameter planetesimals, embryos large enough to accrete gas were barely able to form within 5 My. In the current study, using comparable planetesimals and pebble accretion, multiple gas-giant planets can form within 3 My, and this includes the time required to form planetary embryos and accrete gas. Clearly, the presence of pebble accretion can greatly ease the formation of gas-giant planets within the lifetime of a typical protoplanetary disk.

The values of the pebble fragmentation velocity used here are broadly supported by experimental data for disruptive collisions involving dust aggregates \citep{guttler:2010}. I have ignored the possibility that collisions at lower speeds could lead to bouncing rather than sticking. This could curtail the growth of pebbles at smaller sizes than used here \citep{zsom:2010}. However, some recent work suggests that the sticking properties of space-based materials may be much greater than samples prepared on Earth due to differences in their surface chemistry \citep{schelling:2015}. If so, bouncing collisions may be less important than assumed, and the maximum pebble size could be set by fragmentation rather than bouncing.

The simulations that successfully form giant planets typically have pebbles with diameter $\sim 1$ cm beyond the ice line. This is an order of magnitude larger than most chondrules \citep{friedrich:2015}, which are the principal component of most primitive chondrites. This may mean that the ``pebbles'' used in these simulations actually represent small clusters of chondrules. It has been suggested that such clusters could form due to chondrule-chondrule collisions once chondrules acquired porous, dusty rims that can absorb much of the energy of the impact and increase the chance of sticking \citep{ormel:2008}. Presumably, these chondrule clusters would have rather different collision properties than the materials usually assumed in collision experiments, so the fragmentation velocities used here may not be appropriate. However, the principle that ice-rich particle clusters were larger than rocky ones may still apply.

The model used in this paper does not include planetary migration, which could change the results substantially. One recent study \citep{bitsch:2015} has found that giant planets can form when migration is included at the high rates predicted in theoretical models of migration \citep{tanaka:2002}. Currently there are large uncertainties regarding how and when migration operates \citep{suzuki:2010, duffell:2014, benitez:2015}, but it may be an important factor in planet formation nonetheless. In this paper, I have shown that giant planets can form via pebble accretion under a range of circumstances without migration. Future studies will be needed to see whether this behaviour holds in the presence of migration.

The simulations described in Sections~3 and 4 tend to give a bimodal distribution of outcomes: either several giant planets form rapidly, typically spanning a region of several AU; or no embryos become large enough to become giant planets. However, there are some variations on these two themes. For example, Figure~12 shows the largest embryos formed in 10 simulations at 3 My using a variety of different model parameters. Simulations 1--4 fail to form giant planets, while at least one giant forms in the remaining six cases. As the figure shows, there are significant variations within each of these two subsets. When giant planets do form, for example, the number of giants can vary from a single object to at least eight, although later dynamical interactions will probably reduce this number. In addition, some systems with giants are accompanied by a system of terrestrial planets, while others are not.

The initial radius of the disk clearly plays a role in determining where most of the surviving solid mass is distributed at the end of the simulation. Simulations 3, 4 and 6 use smaller radii (50 or 75 AU) than the other runs, and the surviving material is concentrated closer to the star in each case at 3 My. 

Run 6 is particularly interesting, and unusual, in that a pair of gas-giant planets formed close to the star, far inside the ice line, while only Earth mass planets formed further out. In this case, the high surface density allowed planetesimal-planetesimal collisions near the inner edge of the disk to form bodies massive enough to accrete gas. The low level of turbulence ($\alpha=3\times 10^{-4}$) meant that rocky pebbles grew to about 0.3 cm in diameter in this case, rather than $\sim 0.1$ cm in the cases with larger $\alpha$. These pebbles were still too small for truly effective pebble accretion to occur, and pebbles contributed only about 20 and 30\% of the final solid mass for the two large planets respectively. Pebbles were important in another way, however, since a large amount of mass was transferred to the inner disk in the first 0.1 My in the form of big pebbles undergoing radial drift from the outer disk. This was enough to permit gas-giant cores to form close to the star. In some circumstances at least, giant planets can form without the benefit of pebble accretion dynamics, although pebbles still need to be present to accomplish this.

None of the simulations described here is a perfect match for the planets of the Solar System, but several cases come quite close. For example, Run~5 in Figure~12 ended with a single gas giant planet, and 3 terrestrial planets, together with a belt of objects at larger distances. Two of the terrestrial planets are quite large, roughly 3.5 Earth masses each, so the fit isn't perfect. Presumably, further tinkering with the pebble strength and efficiency of planetesimal formation could yield a closer match, although it isn't clear that this is warranted given other uncertainties in the model. 

Run~7 is also intriguing since it yielded 3 terrestrial planets, each slightly smaller than Earth, plus 4 giant planets, one of which was only 30 times the mass of Earth at 3 My. Beyond the giants lies a disk of small objects, all but one of which is less massive than Earth. Unlike the other simulations in Figure~12, this run began with a disk containing 0.05 solar masses of material---half the mass of the other cases. This suggests that one doesn't necessarily have to adopt the most massive disks to produce giant planets, and somewhat less massive disks could be worth investigating further in future.

One apparent failing of the model used here is the lack of systems with multiple super-Earth-mass planets orbiting close to the star, many examples of which have been found by the Kepler mission \citep{lissauer:2011}. Partly this is a result of the decision to truncate the inner edge of the disk at 0.5 AU. However, this is probably not the whole explanation since super Earths are not seen at larger distances in the simulations, except in the presence of giant planets. It remains unclear what fraction of the Kepler systems contain giant planets orbiting beyond the super Earths, but I can say that the model generally fails to form systems with super Earths alone. This implies that an important piece of physics is still missing from the model. As noted above, orbital migration is one possibility, and this will need to be examined further in future.

%
%
\section{Summary}
In this paper, I have examined the ability of the prevailing model of planet formation to explain the properties of the planets in the Solar System and the diversity of planets discovered in other systems. This model begins with $\mu$m sized dust grains embedded in a gaseous protoplanetary disk. These grains collide and stick to form mm-to-m sized pebbles. Pairwise growth stalls at this point. Large concentrations of pebbles are assumed to form asteroid-sized ``planetesimals'' in a single step via interactions with the gas and self gravity. Planetesimals then collide and gravitationally scatter one another, while also sweeping up pebbles (``pebble accretion''), forming ``planetary embryos''. If embryos grow large enough before the gas disk disperses, they can accrete massive gaseous envelopes, forming gas-giant planets.

The main findings of this study are:
\begin{enumerate}
\item Using typical strengths for rocky pebbles found by experiments, and moderate turbulence levels ($\alpha=7\times10^{-4}$), pebble growth stalls at diameters $\sim 1$ mm. These pebbles are too small to be swept up efficiently by pebble accretion, and planetary growth is very slow. In a massive $0.1$ solar mass disk, the largest objects remain smaller than Mars after 3 My. The increase in solid material beyond the ice line plays a negligible role, and giant planets fail to form.

\item Larger, cm-sized pebbles form beyond the ice line if ice-rich pebbles stick at speeds 3 times higher than rocky pebbles. These pebbles are swept up efficiently by planetesimals and embryos, but the pebbles also drift inwards rapidly due to gas drag. Bodies larger than Earth can form within 3 My, and these are almost massive enough to accrete gaseous envelopes. However, inward drift of pebbles deposits far more mass in the inner disk than exists in the Sun's terrestrial planets.

\item Planetary systems resembling the Solar System can form if (i) icy pebbles are stickier than rocky pebbles, and (ii) the efficiency with which pebbles are converted into planetesimals increases with pebble size. In this case, multiple gas-giant planets form beyond the ice line within 3 My. Pebble accretion is the main factor leading to this rapid growth. Planetesimal formation inside the ice line is relatively inefficient, since pebbles are small here, so terrestrial planets remain small.

\item Most growth occurs early in the disk's lifetime when pebbles are abundant. For the first 0.3 My, the surface density of pebbles inside 5 AU remains high due to the inward drift of pebbles from the outer disk. After 0.3 My, the outer disk becomes depleted in solid material, and the surface density of pebbles declines rapidly everywhere in the disk.

\item The outcome of planet formation depends sensitively on the size of planetesimals that form from pebbles. For efficient pebble accretion, planetesimals must exceed a threshhold mass $\sim 10^{-5}$ Earth masses for cm-sized pebbles. If the largest planetesimals have diameters $<300$ km, they will not reach this threshhold before the supply of pebbles is depleted, and giant planets will fail to form.

\item The outcome does not depend strongly on the planetesimal formation timescale provided that at least a few large planetesimals form while the pebbles remain abundant. Inefficient planetesimal formation may slightly favor giant-planet formation since most of the mass remains in pebbles that can be accreted rapidly by a few large planetesimals.

\item The outcome is sensitive to the level of turbulence in the disk, since this controls pebble-pebble collision speeds and their maximum size. The outcome is also sensitive to the disk radius since this affects the lifetime of pebbles due to radial drift.

\item Density fluctuations in the gas caused by turbulence increase the relative velocities of planetesimals and slow growth. However, this effect is small for the turbulence levels considered here.

\item The sensitivity of pebble accretion to several model parameters typically leads to a dichotomy of outcomes. Either (i) rapid growth via pebble accretion takes place and multiple gas-giant planets form, or (ii) pebble accretion is ineffective and the largest bodies remain comparable to or smaller than Earth. This threshhold effect may explain some of the diversity seen in extrasolar planetary systems.

\end{enumerate}

\acknowledgments
I would like to thank Lindsey Chambers and an anonymous reviewer for helpful comments during the preparation of this manuscript. 

{}

\clearpage
\begin{figure}
\includegraphics[angle=0,scale=.60]{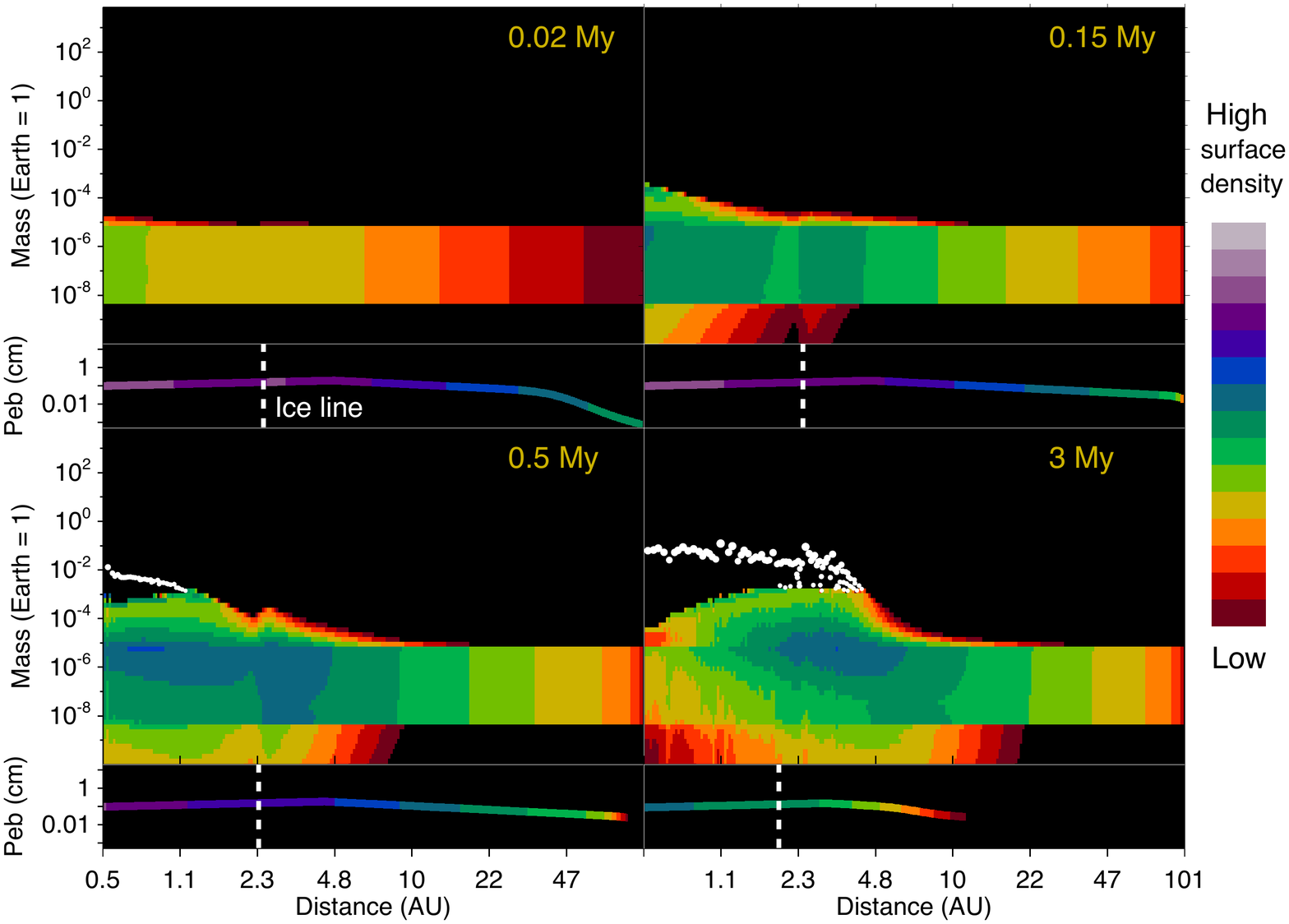}
\caption{The evolution of Case 1 at four snapshots in time. Each of the large panels shows the surface density of planetesimals (colored regions) in each mass bin at each radial location. Different colors denote different surface densities on a log scale, with 3 colors per decade. Planetary embryos are shown on the same plot, indicated by circles. The small panels show the size and surface density of pebbles at each radial location, using the same color scale as the planetesimals. The dashed line shows the location of the ice line (160 K).}
\end{figure}

\begin{figure}
\includegraphics[angle=0,scale=.60]{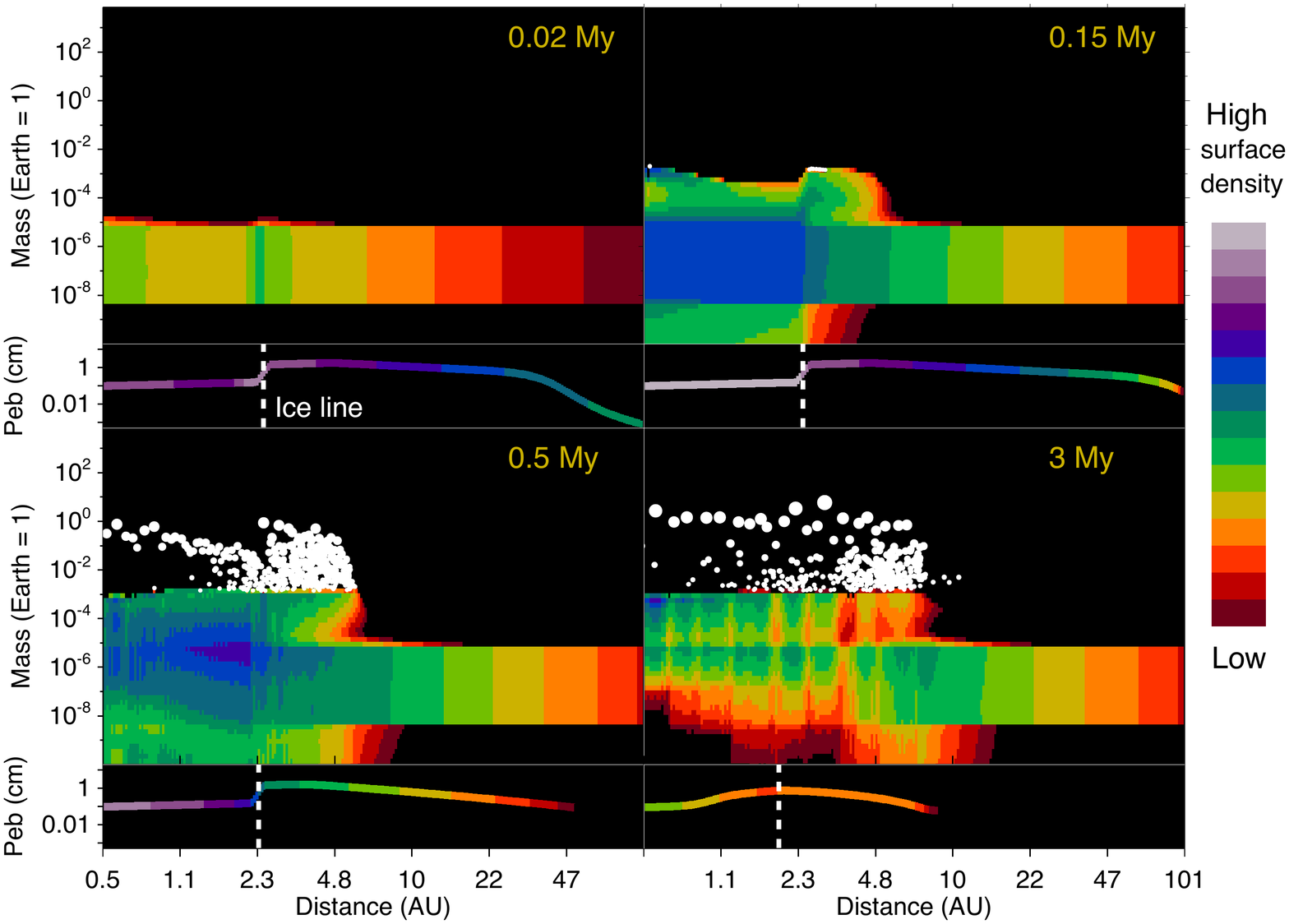}
\caption{The evolution of Case 2 at four snapshots in time. The colors and symbols are the same as those used in Figure~1.}
\end{figure}

\begin{figure}
\includegraphics[angle=0,scale=.60]{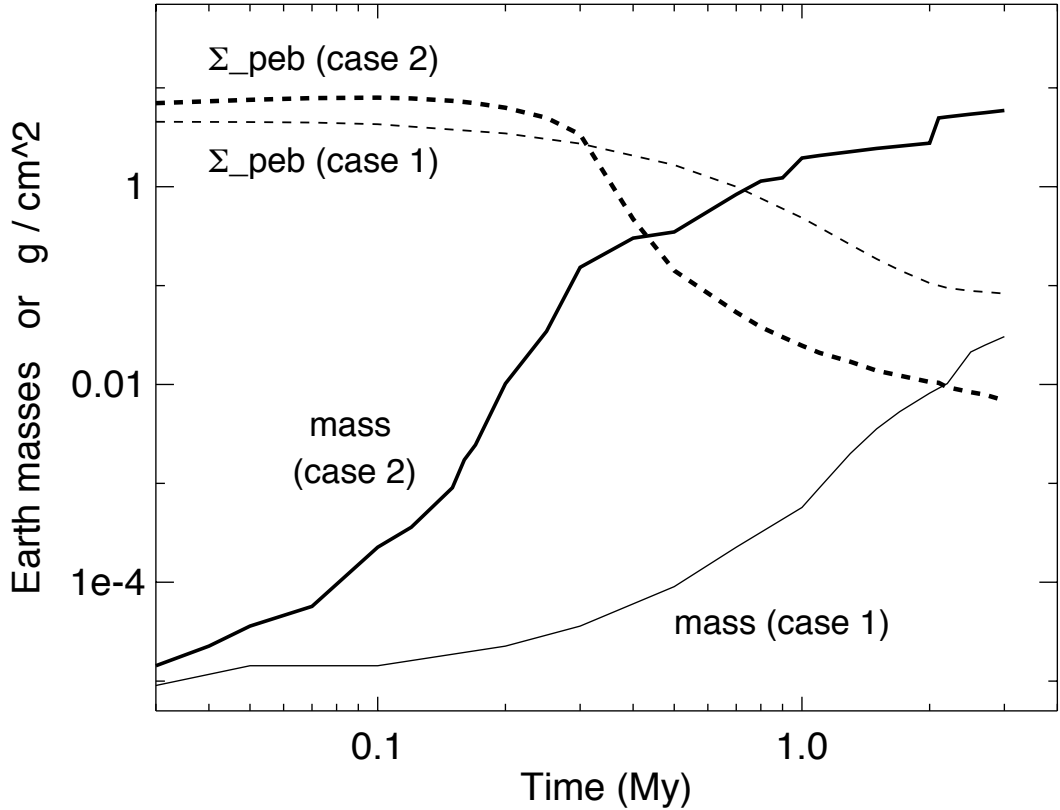}
\caption{Solid curves: the masses of two planetary embryos about 3 AU from the star, one in Case 1 and one in Case 2. (Prior to the appearance of the embryos, the mass of the largest planetesimals at the same location is shown.) Dashed curves: the surface density of pebbles at the same radial location as the embryos.}
\end{figure}

\begin{figure}
\includegraphics[angle=0,scale=.60]{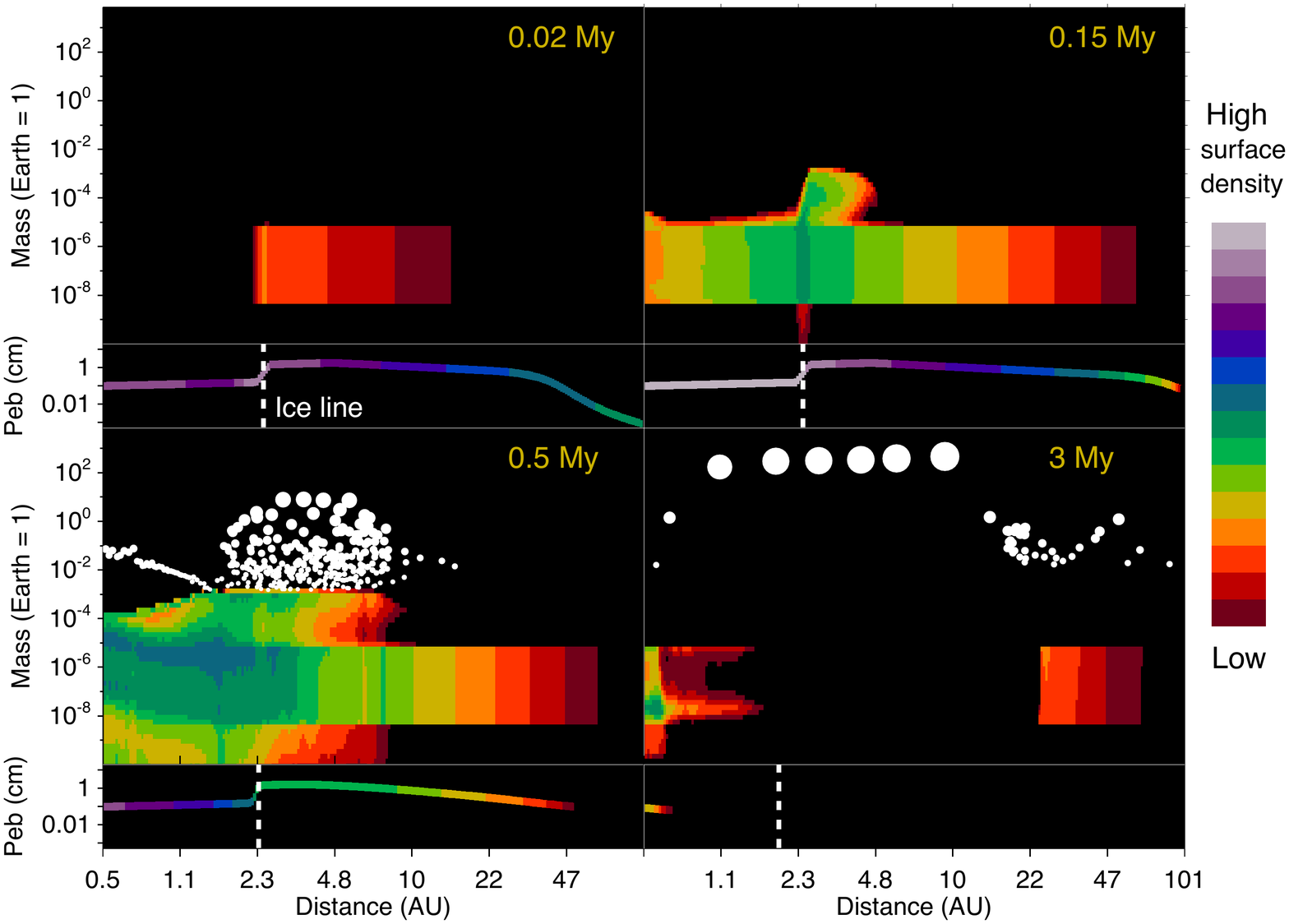}
\caption{The evolution of Case 3 at four snapshots in time. The colors and symbols are the same as those used in Figure~1.}
\end{figure}

\begin{figure}
\includegraphics[angle=0,scale=.60]{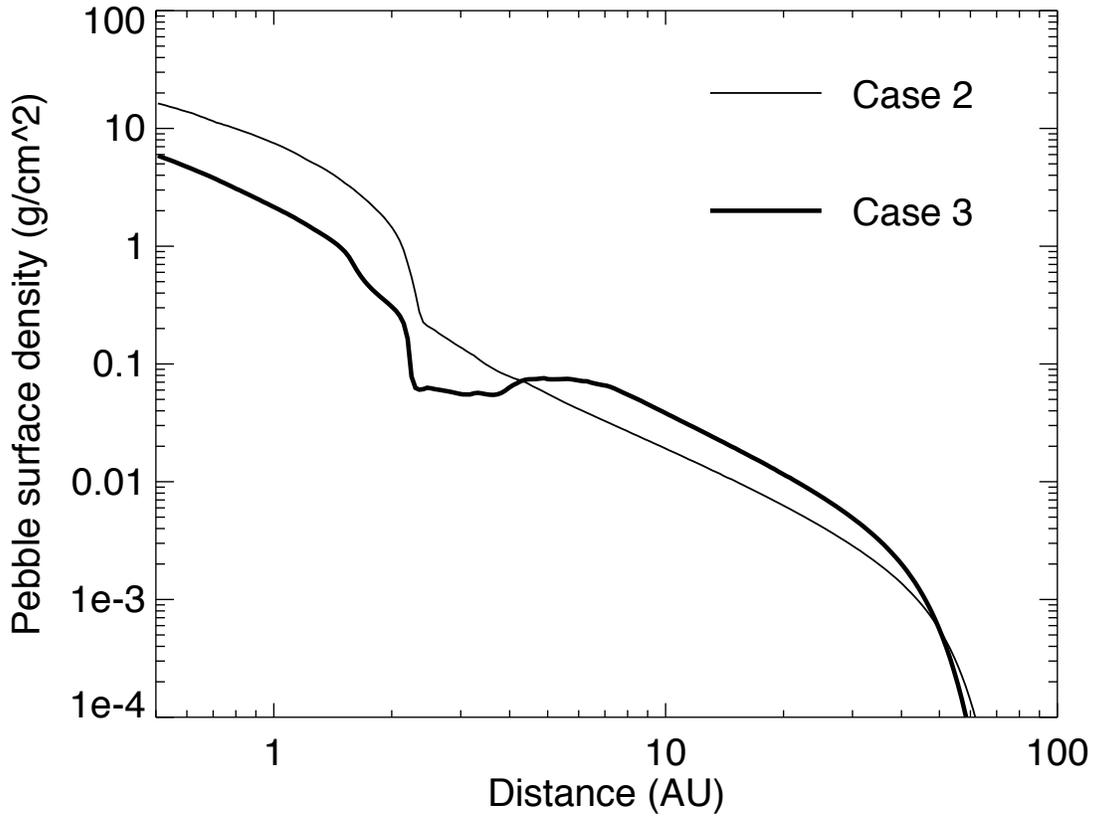}
\caption{The surface density of pebbles versus distance from the star at 0.5 My in Cases~2 and 3.}
\end{figure}

\begin{figure}
\includegraphics[angle=0,scale=.60]{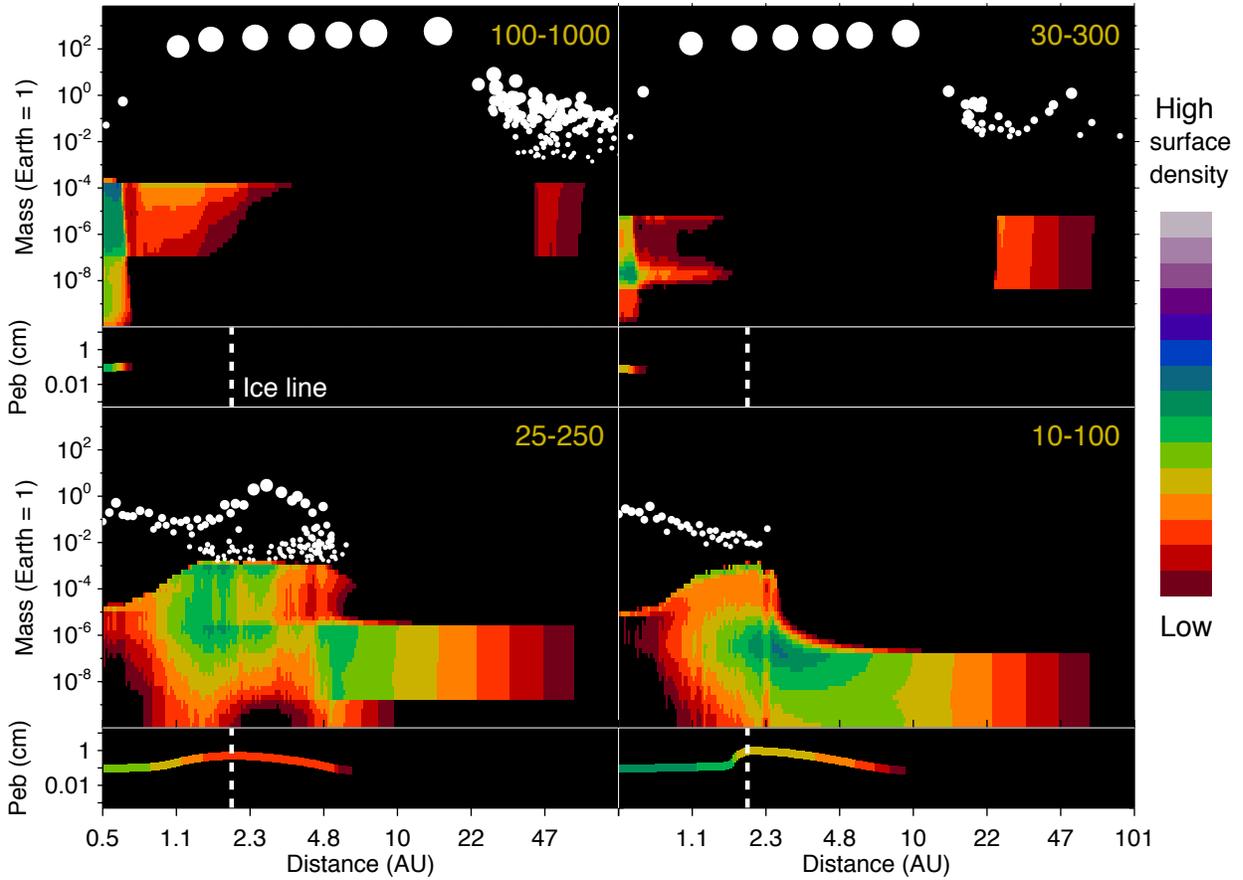}
\caption{The state of four simulations using different initial sizes for planetesimals at 3 My. The range of planetesimal diameters, in km, is indicated in the upper right corner of each large panel. The colors and symbols are the same as those used in Figure~1.}
\end{figure}

\begin{figure}
\includegraphics[angle=0,scale=.60]{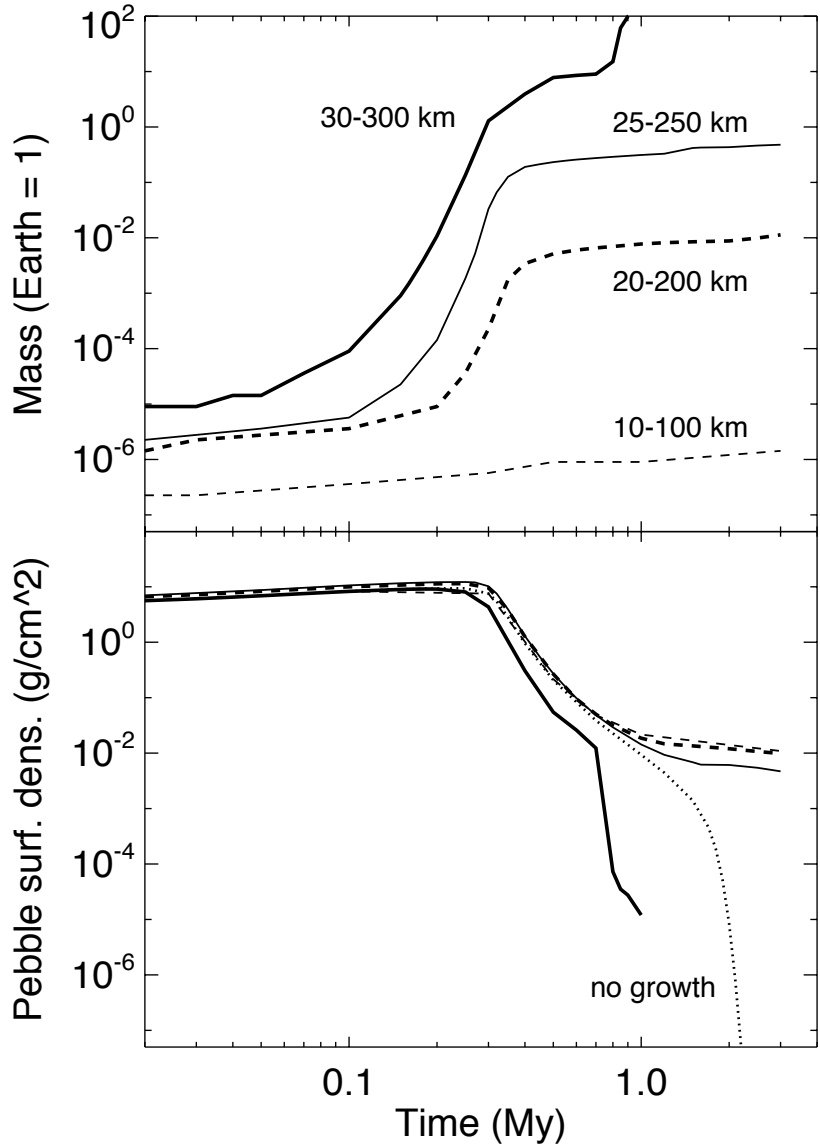}
\caption{Upper panel: the mass of the first large embryo to form outside the ice line versus time for four simulations using different initial planetesimal diameters. Lower panel: the surface density of pebbles at the same radial locations as the embryos versus time. The solid and dashed curves indicate the same planetesimal sizes as the upper panel. The dotted line shows the pebble surface density versus time for a simulation in which planetesimals are prevented from growing after they first form (labelled `no growth').}
\end{figure}

\begin{figure}
\includegraphics[angle=0,scale=.60]{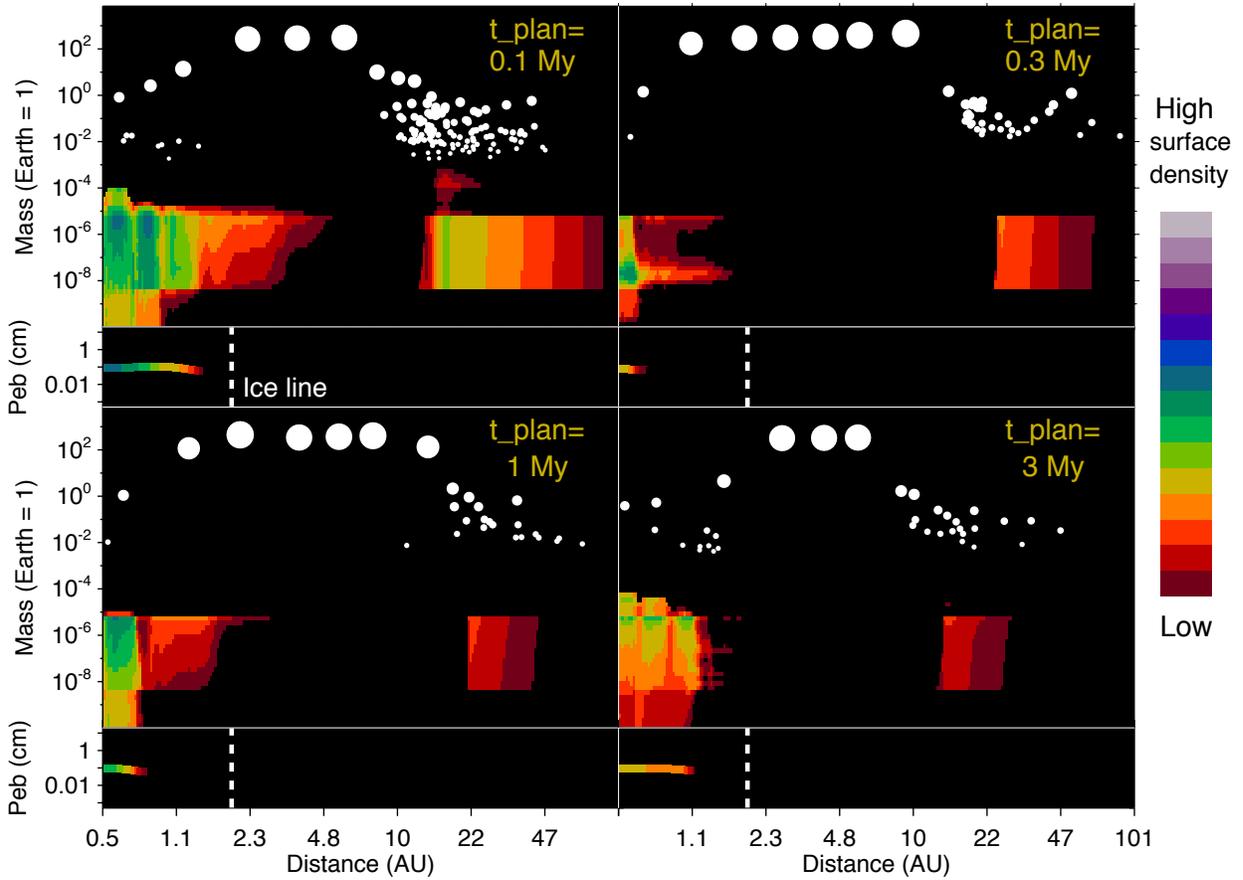}
\caption{The state of four simulations using different planetesimal formation times at 3 My. The planetesimal formation timescale is indicated in the upper right corner of each large panel. The colors and symbols are the same as those used in Figure~1.}
\end{figure}

\begin{figure}
\includegraphics[angle=0,scale=.60]{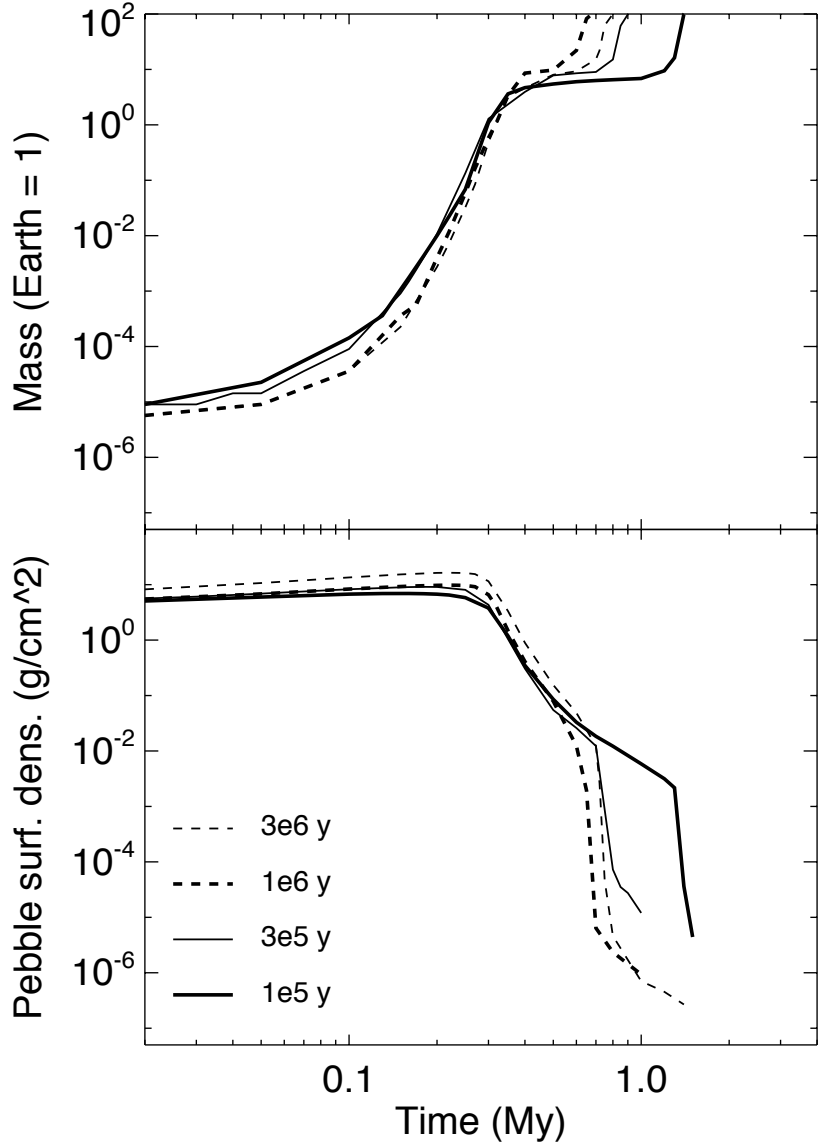}
\caption{Upper panel: the mass of the first large embryo to form outside the ice line versus time for four simulations using different initial planetesimal formation timescales. Lower panel: the surface density of pebbles at the same radial locations as the embryos versus time. The solid and dashed curves indicate the same planetesimal formation timescales as the upper panel.}
\end{figure}

\begin{figure}
\includegraphics[angle=0,scale=.60]{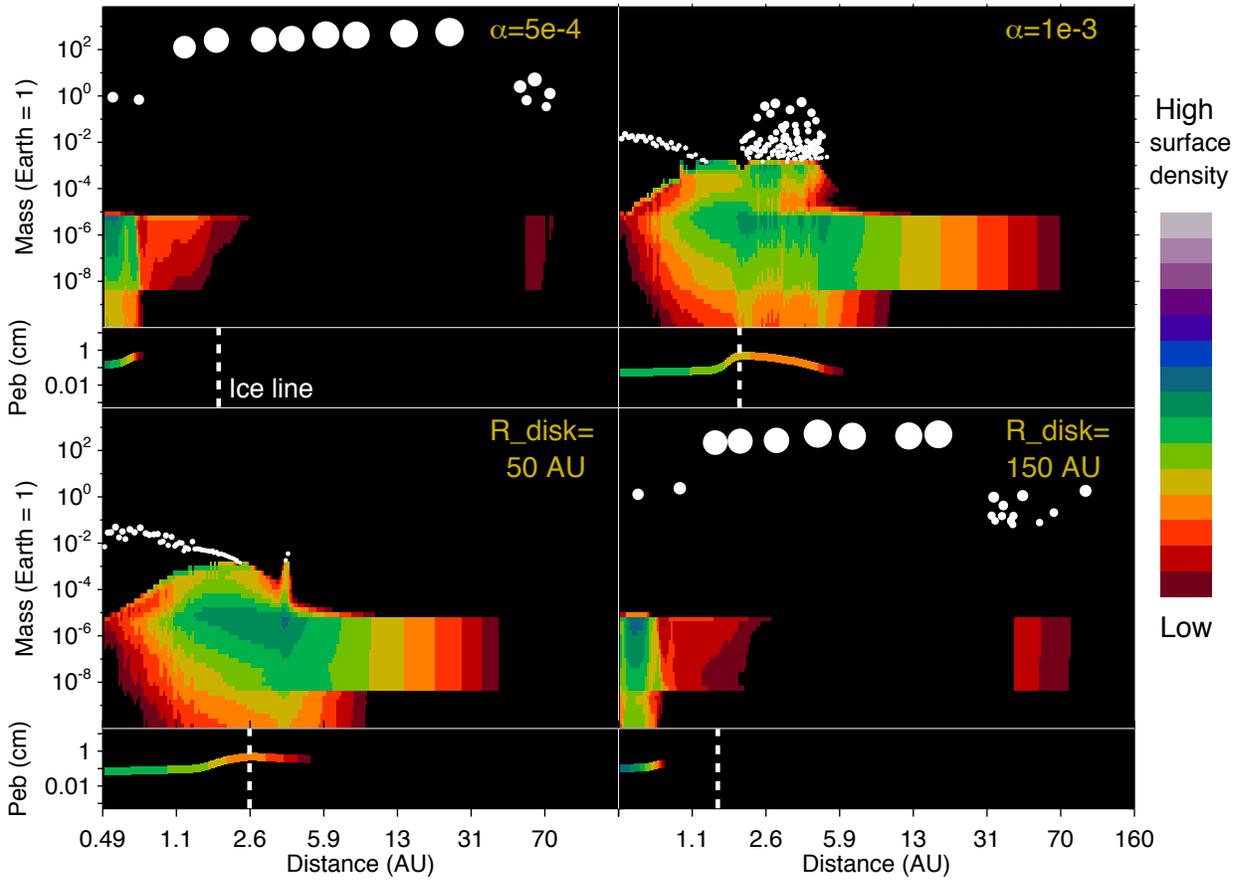}
\caption{The state of four simulations using different disk radii $\rdisk$ and turbulent viscosity parameters $\alpha$ at 3 My. The colors and symbols are the same as those used in Figure~1.}
\end{figure}

\begin{figure}
\includegraphics[angle=0,scale=.60]{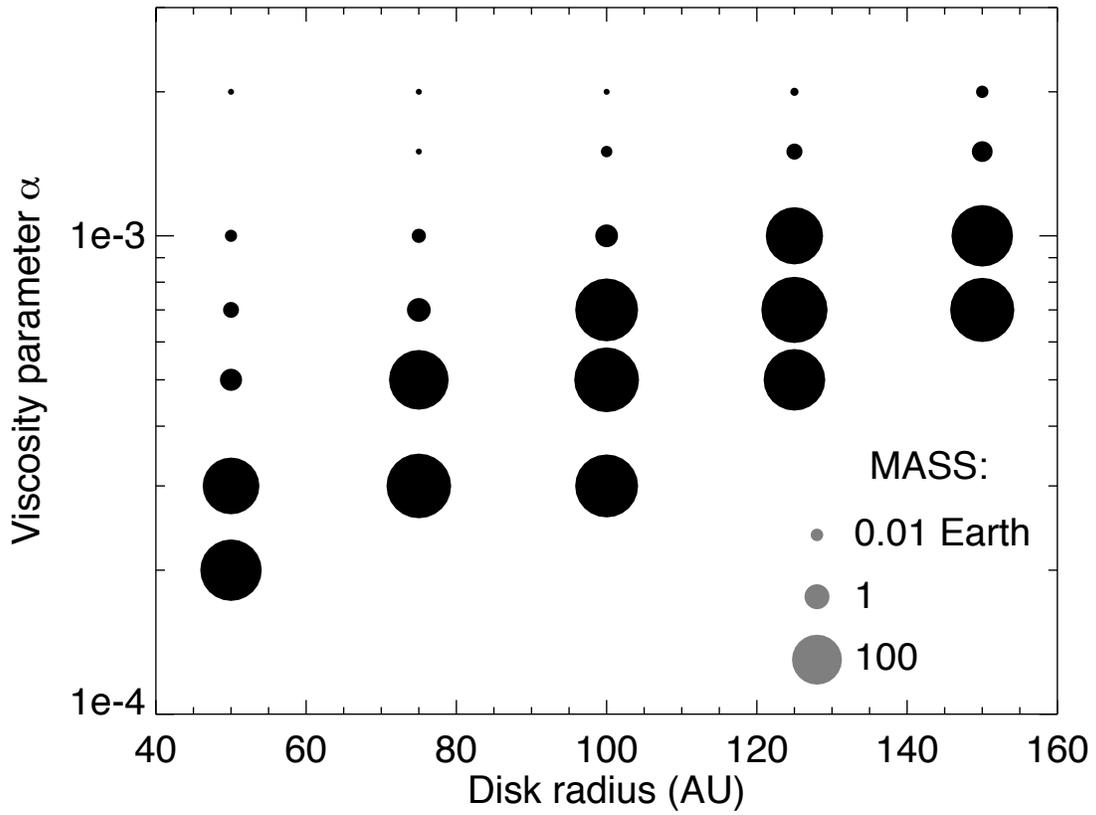}
\caption{The most massive planet to form in each simulation within 3 My for simulations with a range of disk radii and turbulent viscosity parameters.}
\end{figure}

\begin{figure}
\includegraphics[angle=0,scale=.60]{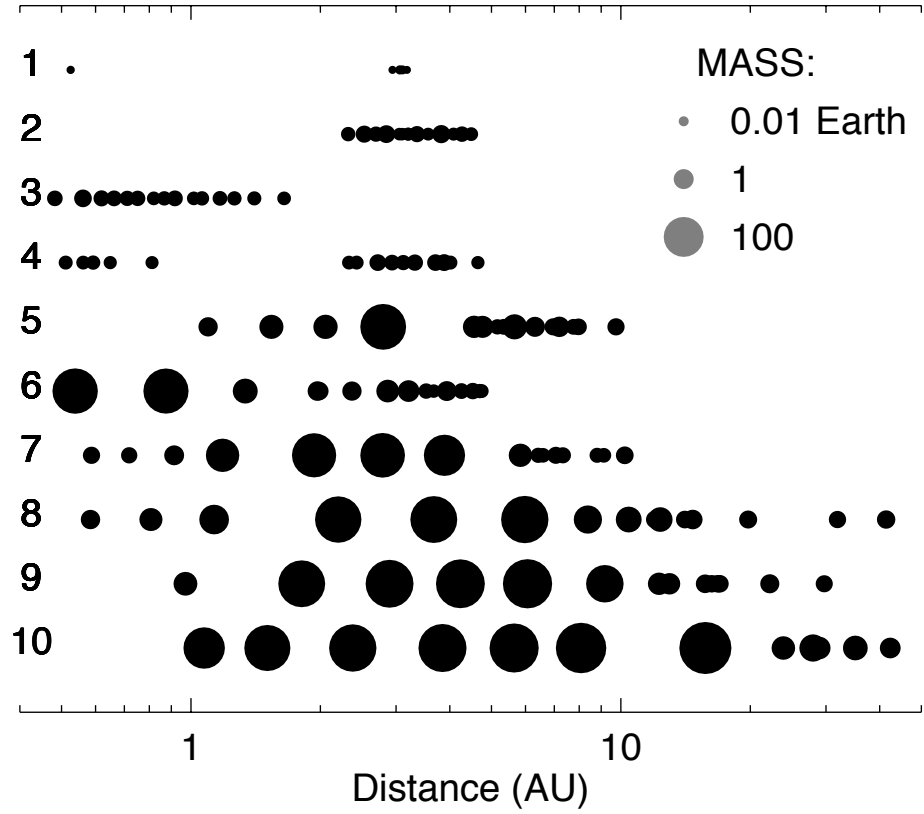}
\caption{The largest objects formed in 10 simulations by 3 My. The parameters used in each simulation are listed in Table~2.}
\end{figure}

\begin{table}
\begin{center}
\begin{tabular}{lcc}
Parameter & Symbol & Default value \\
\hline
Stellar mass & $M_\ast$ & $1\,M_\odot$  \\
Stellar temperature & $T_\ast$ & 4000 K \\
Stellar radius & $R_\ast$ & $3\,R_\odot$ \\
Disk mass & $\mdisk$ & $0.1\,M_\odot$ \\
Disk outer radius & $\rdisk$ & 100 AU \\
Disk inner edge & & 0.5 AU \\
Viscosity parameter & $\alpha$ & $7\times10^{-4}$ \\
Ice to rock mass ratio & & 1:1 \\
Gas to rock mass ratio & & 200:1 \\
Disk gas opacity & $\kappa_{\rm disk}$ & 3 cm$^2$/g \\
Initial pebble diameter & & $1\,\mu$m \\
Pebble fragmentation speed & $\vfrag$ & 1--3 m/s \\
Initial planetesimal diameters & $\dplan$ & 30--300 km \\
Planetesimal formation time & $\tplan$ & 0.3 My \\
Minimum embryo diameter & & 2000 km \\
Planetary atmosphere opacity & $\kappa_{\rm atmos}$ & 0.1 cm$^2$/g \\
Solid bulk density & $\rho$ & 2 g/cm$^2$ \\
Simulation length & $t_{\rm sim}$ & 3 My \\
\end{tabular}
\end{center}
\caption{Main model parameters and default values.}
\end{table}

\begin{table}
\begin{center}
\begin{tabular}{cccccc}
Simulation & $\mdisk$ ($M_\odot$) & $\rdisk$ (AU) & $\alpha$ & $\tplan$ (My) 
& $\dplan$ (km) \\
\hline
1 & 0.1 & 100 & $1.5\times 10^{-3}$ & 0.3 & 30--300 \\
2 & 0.1 & 100 & $1\times 10^{-3}$ & 0.3 & 30--300 \\
3 & 0.1 & 50 & $5\times 10^{-4}$ & 0.3 & 30--300 \\
4 & 0.1 & 75 & $7\times 10^{-4}$ & 0.3 & 30--300 \\
5 & 0.1 & 125 & $1\times 10^{-3}$ & 0.3 & 30--300 \\
6 & 0.1 & 50 & $3\times 10^{-4}$ & 0.3 & 30--300 \\
7 & 0.05 & 100 & $7\times 10^{-4}$ & 0.3 & 30--300 \\
8 & 0.1 & 100 & $7\times 10^{-4}$ & 0.1 & 30--300 \\
9 & 0.1 & 150 & $1\times 10^{-3}$ & 0.3 & 30--300 \\
10 & 0.1 & 100 & $7\times 10^{-4}$ & 0.3 & 100---1000 \\
\end{tabular}
\caption{Model parameters used in Figure~12.}
\end{center}
\end{table}

\end{document}